\documentclass{aa}

\usepackage{graphicx}
\usepackage{natbib}
\bibpunct{(}{)}{;}{a}{}{,} 

\begin{document}

\newcommand{\simgt}{\lower.5ex\hbox{$\;\buildrel>\over\sim\;$}}
\newcommand{\simlt}{\lower.5ex\hbox{$\;\buildrel<\over\sim\;$}}
\newcommand{\hst}{{\sl HST}}
\newcommand{\izw}{I\,Zw\,18}
\newcommand{\sbs}{SBS\,0335$-$052}
\newcommand{\ngc}{NGC\,5253}
\newcommand{\dusty}{{\it DUSTY}}
\newcommand{\dg}{{$\cal D$}}
\newcommand{\ew}{{\sl EW}}
\newcommand{\lya}{{Ly$\alpha$}}                      
\newcommand{\bra}{{Br$\alpha$}}                      
\newcommand{\brg}{{Br$\gamma$}}                      
\newcommand{\ha}{{H$\alpha$}}                      
\newcommand{\hb}{{H$\beta$}}                      
\newcommand{\zsol}{{$Z_\odot$}}                      
\newcommand{\msol}{{$M_\odot$}}                      
\newcommand{\lsol}{{$L_\odot$}}                      
\newcommand{\lir}{{$L_{\rm IR}$}}                      
\newcommand{\tin}{{$T_{\rm in}$}}                      
\newcommand{\tout}{{$T_{\rm out}$}}                      
\newcommand{\rin}{{$R_{\rm in}$}}                      
\newcommand{\rout}{{$R_{\rm out}$}}                      
\newcommand{\yout}{{$Y_{\rm out}$}}                      
\newcommand{\md}{{$M_{\rm dust}$}}                      
\newcommand{\av}{{$A_V$}}
\newcommand{\hii}{{H{\sc ii}}}
\newcommand{\hi}{{H{\sc i}}}
\newcommand{\Ne}{{N$_e$}}
\newcommand{\magsq}{mag\,arcsec$^{-2}$}
\newcommand{\micron}{\,$\mu$m}

\newcommand{\Ho}{\ensuremath{\mathrm{H}_0}}
\newcommand{\Msun}{\ensuremath{~\mathrm{M}_\odot}}
\newcommand{\Lsun}{\ensuremath{~\mathrm{L}_\odot}}
\newcommand{\LBsun}{\ensuremath{~\mathrm{L}_{B\odot}}}
\newcommand{\AV}{\ensuremath{\mathrm{A}_V}}
\newcommand{\QH}{\ensuremath{Q(\mathrm{H})}}
\newcommand{\bh}{\ensuremath{\mathrm{BH}}}
\newcommand{\mbh}{\ensuremath{M_\mathrm{BH}}}
\newcommand{\disp}{\ensuremath{\Delta M_\mathrm{BH}}}
\newcommand{\rbh}{\ensuremath{R_\mathrm{BH}}}
\newcommand{\MB}{\ensuremath{M_\mathrm{B}}}
\newcommand{\MJ}{\ensuremath{M_\mathrm{J}}}
\newcommand{\MH}{\ensuremath{M_\mathrm{H}}}
\newcommand{\MK}{\ensuremath{M_\mathrm{K}}}
\newcommand{\msph}{\ensuremath{M_\mathrm{bul}}}
\newcommand{\mdyn}{\ensuremath{M_\mathrm{dyn}}}
\newcommand{\lsph}{\ensuremath{L_\mathrm{bul}}}
\newcommand{\LV}{\ensuremath{L_\mathrm{V}}}
\newcommand{\lBsph}{\ensuremath{L_\mathrm{B,bul}}}
\newcommand{\lRsph}{\ensuremath{L_\mathrm{R,bul}}}
\newcommand{\lJsph}{\ensuremath{L_\mathrm{J,bul}}}
\newcommand{\lHsph}{\ensuremath{L_\mathrm{H,bul}}}
\newcommand{\lKsph}{\ensuremath{L_\mathrm{K,bul}}}
\newcommand{\lnirsph}{\ensuremath{L_\mathrm{NIR,bul}}}
\newcommand{\lopt}{\ensuremath{L_\mathrm{opt}}}
\newcommand{\sigstar}{\ensuremath{\sigma_\mathrm{e}}}
\newcommand{\re}{\ensuremath{R_\mathrm{e}}}

\newcommand{\ten}[1]{\ensuremath{10^{#1}}}
\newcommand{\xten}[1]{\ensuremath{\times 10^{#1}}}
\newcommand{\1}{\ensuremath{^{-1}}}
\newcommand{\2}{\ensuremath{^{-2}}}
\newcommand{\3}{\ensuremath{^{-3}}}

\newcommand{\CM}{\ensuremath{\mathrm{~cm}}}
\newcommand{\KM}{\ensuremath{\mathrm{~km}}}
\newcommand{\PC}{\ensuremath{\mathrm{~pc}}}
\newcommand{\KPC}{\ensuremath{\mathrm{~kpc}}}
\newcommand{\MPC}{\ensuremath{\mathrm{~Mpc}}}

\newcommand{\SEC}{\ensuremath{\mathrm{~s}}}
\newcommand{\YR}{\ensuremath{\mathrm{~yr}}}

\newcommand{\kms}{\KM\SEC\1}
\newcommand{\ERG}{\ensuremath{\mathrm{~erg}}}
\newcommand{\ARCSEC}{\ensuremath{\mathrm{~arcsec}}}

\newcommand{\vel}{\ensuremath{\langle v\rangle}}
\newcommand{\vcirc}{\ensuremath{v_\mathrm{circ}}}
\newcommand{\velsq}{\ensuremath{\langle v^2\rangle}}

\newcommand{\HA}{\ensuremath{\mathrm{H}\alpha}}
\newcommand{\HB}{\ensuremath{\mathrm{H}\beta}}
\newcommand{\NII}{\ensuremath{\mathrm{[N\,II]}}}
\newcommand{\OIII}{\ensuremath{\mathrm{[O\,III]}}}
\newcommand{\OII}{\ensuremath{\mathrm{[O\,II]}}}
\newcommand{\SII}{\ensuremath{\mathrm{[S\,II]}}}

\newcommand{\mlr}{\ensuremath{\Upsilon}}
\newcommand{\I}{\ensuremath{i}}
\newcommand{\Th}{\ensuremath{\theta}}
\newcommand{\B}{\ensuremath{b}}
\newcommand{\So}{\ensuremath{\mathrm{s}_\circ}}
\newcommand{\Vsys}{\ensuremath{V_\mathrm{sys}}}
\newcommand{\chisq}{\ensuremath{\chi^2}}
\newcommand{\chisqr}{\ensuremath{\chi^2_\mathrm{red}}}
\newcommand{\chisqc}{\ensuremath{\chi^2_\mathrm{c}}}

\title{The optical-to-radio spectral energy distributions of low-metallicity
blue compact dwarf galaxies}

\author{Leslie Hunt \inst{1},
   Simone Bianchi \inst{1},
\and
   Roberto Maiolino \inst{2}
   }

\offprints{L. K. Hunt, \email{hunt@arcetri.astro.it}}

\institute{INAF-Istituto di Radioastronomia/Sez. Firenze, 
	Largo Enrico Fermi, 5, 50125 Firenze, Italy\\
       \email{hunt@arcetri.astro.it,sbianchi@arcetri.astro.it}
  \and
       INAF - Osservatorio Astrofisico di Arcetri, Largo Enrico Fermi, 
       5, 50125, Firenze, Italy\\
       \email{maiolino@arcetri.astro.it}
      }

\date{Received ; accepted}

\abstract{ 
We present 
global spectral energy distributions (SEDs) from 0.3\,$\mu$m to 90\,cm
for a sample of seven low-metallicity blue compact dwarf galaxies (BCDs).
In addition to data compiled from the literature, we report new SCUBA data
for the galaxies in the sample, including the two 
most metal-poor star-forming galaxies known, \izw\ and \sbs.
The standard starburst templates, M\,82 and Arp\,220, do not
give a good approximation to any of the BCD SEDs in our sample.
Several SEDs are instead characterized by: {\it (i)}~approximately flat radio 
spectra suggesting
dominant thermal processes; {\it (ii)}~far-infrared (FIR) spectra which peak at or
shortward of 60\,$\mu$m; {\it (iii)}~a significant warm dust component
and absence of Aromatic Features in Emission (or PAHs) in the mid-infrared (MIR).
The SEDs of low-metallicity BCDs do not follow ``canonical'' scaling relations,
and the ``standard'' star-formation indicators based on radio continuum,
FIR, and MIR luminosities can be discrepant by factors of $\simgt10$.
We present new models of the dust SEDs from 5\,$\mu$m to 1\,cm, and derive
infrared luminosities, dust distributions, temperatures, and masses.
The observed SEDs and dust models are interpreted
in terms of the active/passive regimes of
star formation and trends with metallicity. Finally, we discuss the 
implications of our results on the $z\simgt6$ starburst populations which will
be detected by forthcoming submm and radio facilities.
\keywords{Galaxies:starburst -- Galaxies:compact --
Infrared:galaxies -- Radio continuum:galaxies }}

\authorrunning{Hunt, Bianchi, \& Maiolino}
\titlerunning{SEDs of low-metallicity BCDs}

\maketitle

\section{Introduction}

The physical processes which governed the formation of the first stars 
in the universe and the epoch during which it occurred
are a main focus of modern cosmology.
Measuring the cosmic star formation rate
(SFR) of the universe as a function of redshift or look-back time
has been the subject of much effort,
initially at ultraviolet wavelengths \citep[e.g.,][]{madau},
and more recently in the millimeter and radio regimes
\citep[e.g.,][]{blain99,haarsma}.
Unlike the ultraviolet or optical spectral regions,
emission at radio and (sub) millimeter wavelengths is unaffected by dust,
which makes uncertain extinction corrections unnecessary.
To exploit this advantage,
emission at virtually every wavelength from the mid-infrared (MIR) to the radio
has been used to trace star formation, both locally 
\citep{kennicutt98,roussel01,condon91}
and at $z\simlt4$ \citep{rr97,blain99,carilli,chary,garrett,yun}.
Nevertheless, the conversion of MIR, far-infrared (FIR), 
sub-mm or radio continuum flux to a global
SFR depends on canonical scaling relations (radio/FIR, MIR/FIR),
which may not be applicable to primordial star-forming systems 
at $z\geq5$. 

Even locally,
there are some indications that SFR is not adequately traced by FIR
or radio luminosity of low-luminosity or young star-forming galaxies 
\citep{klein91,roussel03}.
Indeed, in dwarf galaxies, the scaling relations appear to hold 
only because both IR and radio emission underestimate the
true SFR \citep{bell03}.
In these objects, particularly those of low metal abundance,
the SFR estimated assuming the standard radio mix of thermal/non-thermal
contributions can be underestimated by a factor of 5 or more
\citep{kj,beck02,hunt04a}. 
While rare in the nearby universe \citep[see ][]{kunth}, low-metallicity low-mass
galaxies might be much more frequent at early times and high
redshifts, given the predictions of the
hierarchical merger models \citep[e.g.,][]{baugh98,cole}.
Indeed, such objects may represent the primordial ``building blocks'' --
or ``sub-galaxies'' \citep{rees} --
in hierarchical scenarios of galaxy formation.

To better understand the effects of metallicity and luminosity
on the properties of star-forming regions and investigate possible 
spectral templates
for high-$z$ galaxies, we have undertaken a 
study of the spectral energy distributions
(SEDs) of low-metallicity blue compact dwarfs (BCDs).
Low-metallicity galaxies are in fact the most viable local laboratories
with which to study primordial galaxy formation,
because of their low mass and chemically unenriched interstellar medium (ISM).
However,
such objects cannot be compared to the massive ``SCUBA galaxies'' and Lyman-break
systems which are already chemically enriched at $z\simlt3$
\citep{pettini01,tecza,shapley04}.
At higher redshifts $z\simgt6$,
truly primordial star formation must occur in very low or
zero metallicity environments, and metal-poor BCDs are the only way to study
such processes locally.
Because IR--radio correlations are used to infer SFR for
galaxies observed at high redshift, it is important to verify
that all galaxies, even the most chemically unevolved, follow
such correlations.
Moreover, dust grain properties may depend on metallicity
and on the age of the stellar populations.
Stellar ages at cosmological distances have a strong bearing on the origin
of the dust (stellar winds of evolved stars versus supernovae).

In this paper,
we present global fluxes from 0.3\micron\ to 90\,cm
for 7 BCDs with metallicities ranging from 1/50 \zsol\ to solar.
In addition to our new near-infrared (NIR) images, we have
also acquired data from the HST and JCMT/SCUBA archives, and gleaned 
optical, infrared, and radio data from the literature.
In Sect. \ref{sec:sampledata}, we describe the sample, the data,
and the data reduction.
Sect. \ref{sec:seds} presents the SEDs for the sample BCDs,
and Sect. \ref{sec:active} discusses their interpretation in terms
of ``active'' and ``passive'' star formation.
New SED models for the BCDs in our sample with sufficient data
are given in Sect. \ref{sec:dusty}, 
and scaling laws among mid-, far-infrared, and radio emission
are described in Sect. \ref{sec:scaling}.
We discuss the implications of our results on star-formation
history and metallicity in Sects. \ref{sec:age} and \ref{sec:metals}, and
in the context of high-redshift star 
formation in Sect. \ref{sec:highz}.
If low-metallicity BCD SEDs differ from the usual templates 
(e.g., M\,82 and Arp\,220), it will be necessary to reassess
the interpretation of
the faint sources at $z\simgt6$ expected to be detected with the new
sub-mm, mm, and radio facilities such as APEX, ALMA, SKA, and LOFAR.

\section{The sample and the data \label{sec:sampledata}}

Our sample is based on all the low-metallicity
BCDs available from the JCMT/SCUBA archive as of December, 2003. 
This resulted in seven BCDs with metallicities ranging from
solar  to \simlt\,1/50 \zsol.
Our sample includes the two most metal-poor star-forming galaxies
known, \izw\ (\simlt\,1/50 \zsol) and \sbs\ (1/41 \zsol).
Table \ref{tab:sample} lists the basic properties of the sample.
Properties are taken from NED, with the exception of the metallicities,
which were gleaned from the references listed in the last column of
the Table.
Distances were calculated assuming a Hubble constant 
$H_0\,=\,70$\,km\,s$^{-1}$\,Mpc$^{-1}$, after correcting the heliocentric
radial velocities to the CMB reference frame according to the prescription
of the Third Reference Catalogue (RC3, de Vaucouleurs et al. 1991).
While this sample is neither complete nor homogeneously selected, it contains
many heavily studied BCDs and enables a comparison of their spectral properties.
It also has the virtue of spanning a factor of 50 in metallicity,
and a factor of $\simgt 10^4$ in blue luminosity.
With the exception of Mrk\,33,
all the sample galaxies are BCDs according to the criteria (luminosity,
central color, peak surface brightness) set out by \citet{gildepaz}.
Mrk\,33, with an absolute $B$ magnitude of $-18.4$, is slightly too luminous
to be considered a true dwarf galaxy.

\begin{table*}
\caption[]{Sample galaxies \label{tab:sample}}
\begin{tabular}{lrrrcccrrl}
\multicolumn{1}{c}{Name} &
\multicolumn{1}{c}{$a$} &
\multicolumn{1}{c}{$b$} &
\multicolumn{1}{c}{$z$} &
\multicolumn{1}{c}{D (Mpc)$^a$} &
\multicolumn{1}{c}{M$_B$} &
\multicolumn{1}{c}{$12+\log$(O/H)} &
\multicolumn{1}{c}{\zsol/$Z^b$} &
\multicolumn{1}{c}{A$_B^c$} &
\multicolumn{1}{c}{Ref.$^d$}\\
\hline
\izw		& 0.4 & 0.2 & 0.0025 & 13.0  & -14.1  & 7.19 & 55 & 0.138 & 1 \\
\sbs		& 0.2 & 0.2 & 0.0135 & 55.7  & -16.7  & 7.32 & 41 & 0.202 & 1 \\
II\,Zw\,40 	& 0.6 & 0.2 & 0.0026 & 12.1  & -14.9  & 8.09 &  7 & 3.538 & 2,3 \\
II\,Zw\,70=UGC\,9560	& 0.7 & 0.3 & 0.0041 & 19.5  & -16.7 &  8.06 & 7 & 0.053 & 3 \\
\ngc	        & 5.0 & 1.9 & 0.0014 &  4.1$^e$ & -17.2  & 8.19 & 5 & 0.242 & 1,3,4 \\
Mrk\,33=Haro\,2	& 1.0 & 0.9 & 0.0049 & 23.4  & -18.4  & 8.40 & 3 & 0.052 & 5 \\
He\,2-10	& 1.9 & 1.4 & 0.0029 &  9.0$^f$  & -17.3  & 8.93 & 1 & 0.481 & 4 \\
\hline
\end{tabular}

\smallskip
$^a$Distances calculated with $H_0\,=\,70$\,km\,s$^{-1}$\,Mpc$^{-1}$,
and correcting to the CMB reference frame as described in the text.
The distance for \izw\ derived by \citet{ostlin} is 12.6\,Mpc. \\
$^b$Assuming solar $12+\log$O/H\,=\,8.93 \citep{anders}. \\
$^c$Galactic extinction $A_B$ (mag) taken from \citet{schlegel}.\\
$^d$Reference for metallicity: 
(1)~\citet{it99};
(2)~\citet{guseva};
(3)~\citet{ks};
(4)~\citet{kobulnicky};
(5)~\citet{legrand}. \\
$^e$Distance taken the Cepheid variable estimate given by \citet{sandage94}. \\
$^f$Distance taken from \citet{kj}.\\
\end{table*}

\subsection{SCUBA data reduction\label{sec:scuba}}

Data reduction of SCUBA data was carried on in the standard way,
using the dedicated software SURF by \citet{jenness1998}. The data
were first flat-fielded to take into account different bolometer
sensitivities, then corrected for atmospheric extinction. 
Atmospheric opacities were mainly derived from the 225GHz opacity 
monitor, using the well studied relations between the optical 
depth at 225GHz and those at 450$\mu$m and 850$\mu$m. Bolometers 
that appeared free of source emission were used to remove residual 
sky noise fluctuations. Flux calibration was achieved from observations 
of secondary calibrators or Uranus, when available.

Five objects were observed in jiggle-map mode, with data available
for both the short (450$\mu$m) and long  (850$\mu$m) wavelength 
arrays. Maps were produced after rebinning the data in the RA-Dec 
plane. Total fluxes were measured integrating
the source signal (and the calibrator) over a selected aperture,
as described in \citet{dunne2000} and \citet{jenness2002}. For
detected sources, apertures were selected to enclose regions where 
S/N$>$2 (unless they were smaller than a 40'' circular aperture,
which was used in these cases). II\,Zw\,40 was observed in 10 nights 
between February 2000 and January 2001, for a total number of 315 
integrations (10 integrations require about 20 minutes to be completed, 
with the bolometer being on-source half of the time). 
The sky opacity at 850$\mu$m was in the range $\tau_{850} =0.2-0.3$. 
The object was detected both at 450$\mu$m and 850$\mu$m. At the 
longer wavelength the galaxy is well resolved by SCUBA (FWHM$\approx$15'').
He\,2$-$10 was observed in two nights in 
December 2000 (95 integrations) with sky opacity $\tau_{850} =0.2-0.3$.
The object has also been detected at both wavelengths.
II\,Zw\,70 and \ngc\ were observed together in a single night 
(1999, January 13) with $\tau_{850} =0.3-0.4$ (20 integrations each).
Only \ngc\ was detected, at 850$\mu$m. Finally, Mrk\,33
was observed during several nights between March 1998 and January 1999
(106 integrations), with $\tau_{850}$ in the range $0.1-0.5$. It also has
been detected at 850$\mu$m. Fluxes for these objects (and 2-$\sigma$
upper limits) are reported in Tables \ref{tab:scuba}. To our
knowledge, the data for only two of these objects have been previously
published \citep[namely \ngc\ and Mrk\,33][]{james02}; the fluxes
from the literature are in excellent agreement with our independent data
reduction, the results of which are reported in Table \ref{tab:scuba}.

The two remaining objects of our sample, \izw\ and \sbs\, have been
observed in photometry mode in February 2000, in nights with optical
depths $\tau_{850}=0.3-0.4$. A total number of 497 integrations were
spent on \izw\, and 200 on \sbs. Neither object was detected; upper
limits are given in Tables \ref{tab:scuba}.

\begin{table*}
\caption[]{Total SCUBA fluxes for sample galaxies \label{tab:scuba}}
\begin{tabular}{lrrl}
\multicolumn{1}{c}{Name} &
\multicolumn{1}{c}{450$\mu$m} &
\multicolumn{1}{c}{850$\mu$m} &
\multicolumn{1}{c}{Comments}\\
&\multicolumn{1}{c}{(mJy)} & \multicolumn{1}{c}{(mJy)} \\
\hline
\izw		&\multicolumn{1}{c}{--} & $<$ 2.5 \\
II\,Zw\,40 	& 1550$\,\pm$\,350 & 90\,$\pm$\,10 & Resolved at 850$\mu$m \\
\sbs		&\multicolumn{1}{c}{--} & $<$ 5 \\
II\,Zw\,70=UGC\,9560	&\multicolumn{1}{c}{--} & $<$ 20 \\
\ngc	        & $<$2200 & 180\,$\pm$\,20 & Resolved at 850$\mu$m \\
He\,2-10	& 580$\,\pm$\,150 & 140$\,\pm$\,15 & Resolved at 850$\mu$m \\
Mrk\,33=Haro\,2	& $<$170 & 50$\,\pm$\,10 & Resolved at 850$\mu$m\\
\hline
\end{tabular}
\end{table*}

\subsection{Data from the literature \label{sec:data}}

We searched the literature for optical, NIR, mid- and far-infrared,
sub-mm, mm, and radio data.
Data from our own observations were incorporated for
Mrk\,33 (Haro\,2), II\,Zw\,40, and \sbs.
Because we aim in this paper to derive the SED for the entire galaxy,
rather than individual regions, we have placed
greater emphasis on the largest aperture available and total fluxes. 
This is a problem particularly for the older
mid-infrared (MIR) observations with photometers, 
which were, of necessity, performed with small apertures.
Global fluxes are frequently
unavailable, and comparison of different aperture sizes
at different wavelengths is
the major source of the uncertainties in our SEDs.

For wavelengths $<$ 5\micron,
Galactic extinction was corrected according to the values given by
\citet{schlegel} (see Table \ref{tab:sample}), and using the Galactic
extinction curve with $R_V=3.1$ by \citet{cardelli}.
Corrected optical magnitudes were converted to fluxes using the zero points by
\citet{bessell79}, and NIR/MIR ones given by the UKIRT web 
page\footnote{{\tt www.jach.hawaii.edu/JACpublic/UKIRT/astronomy/} \hfill\\
{\tt conver.html}}. 
The UKIRT web page is based on the calibrations given by
\citet{beckwith} and \citet{tokunaga}.
Interpolation was applied as necessary.

Fluxes from the {\it Infrared Astronomical Satellite} (IRAS) were 
taken either from the NASA/IPAC Extragalactic Database (NED), or 
from published papers. 
Although data were examined on a case-by-case basis,
single-dish radio observations were generally preferred to 
high-resolution VLA data, because of the possibility of losing
flux from low spatial frequency filtering.

\subsection{ISO data for \sbs \label{sec:iso}}

To re-evaluate the 65\,$\mu$m flux for \sbs, 
in light of the Spitzer\footnote{The Spitzer Space Telescope is a NASA mission 
managed by the Jet Propulsion Laboratory. 
See {\tt www.spitzer.caltech.edu}.}
results \citep{houck04},
we extracted from the ISO archive the ISOPHOT data for this object.
The pipeline reduction and calibration was used, and aperture photometry was performed
with IRAF\footnote{IRAF is the Image Analysis and 
Reduction Facility made available to the astronomical community by the National Optical
Astronomy Observatory, which is operated by AURA, Inc., under
contract with the U.S. National Science Foundation.}.
We obtain a weak ($\sim\,2\sigma$) detection for \sbs\ at 65\,$\mu$m of $\sim$44\,mJy.
This is substantially lower than the 112\,mJy estimate by \citet{ps},
but in much better agreement with the Spitzer IRS spectrum
reported by \citet{houck04}.

\subsection{NIR data for Mrk\,33 and II\,Zw\,40 \label{sec:nir}}

For two BCDs in our sample, we present new NIR imaging data.
$J$ (1.2\micron), $H$ (1.6\micron), and $K$ (2.2\micron)
images of II\,Zw\,40 were acquired with the 3.8-m United Kingdom Infrared Telescope
(UKIRT\footnote{The United Kingdom Infrared Telescope is operated by the
Joint Astronomy Centre on behalf of the U.K. Particle Physics
and Astronomy Research Council.})
equipped with IRCAM3.
The observations are part of our ongoing project of NIR imaging and
spectroscopy of BCDs \citep[see][]{hunt03}.
The IRCAM3 plate scale is 0\farcs28 per pixel, 
with a total field-of-view (FOV) of 72\arcsec~$\times$~72\arcsec.
We acquired $J$, $H$, and $K$ images of Mrk\,33 with ARNICA mounted on
the 1.5-m Italian Telescope at Gornergrat 
(TIRGO\footnote{TIRGO (Gornergrat, CH) is operated by the 
Istituto di Radioastronomia-Sezione Firenze.}).
The ARNICA plate scale is 0\farcs97 per pixel with a FOV of
4.1$^\prime$$\times$4.1$^\prime$.
Both cameras house NICMOS3 256$^2$ detectors.

Each galaxy was observed by alternating frames on the source and on
adjacent empty sky positions,
beginning and ending each observing sequence with a sky position.
For the UKIRT data,
before the beginning of each sequence,
dark exposures were acquired with the same parameters as the subsequent
science frames.
Individual frames were dark-subtracted (in the case of IRCAM3)
and flat-fielded with 
the average of adjacent empty sky frames,
after editing them for stars (to avoid ``holes'' in the reduced frames)
and applying an average-sigma clipping algorithm.
The reduced frames were then aligned and averaged.
All data reduction was carried out in the 
IRAF environment\footnote{IRAF is the Image Analysis and Reduction Facility
made available to the astronomical community by the National Optical
Astronomy Observatory, which is operated by AURA, Inc., under
contract with the U.S. National Science Foundation.}.

Photometric calibration at both telescopes was performed by observing standard stars
from the UKIRT Faint Standard List \citep{hawarden}
before and after the source observations.
Each standard star was measured in several different positions on the array,
and flat-fielded by dividing the clipped mean of the remaining frames 
in the sequence. To correct the standard-star 
photometry for atmospheric extinction,
we used the mean extinction coefficients given by each observatory.
Virtual aperture photometry was performed for both objects, but Mrk\,33 is
sufficiently large that we opted for
asymptotic magnitudes derived from growth curves, rather than
truncating to a given aperture.
The magnitudes derived in this way agree to within 0.1\,mag to the total
magnitudes given by \citet{jarrett}.
Galactic extinction was corrected for and conversion to flux
was achieved as described in Sect. \ref{sec:data}.

\section{Spectral energy distributions \label{sec:seds}}

The SEDs of our sample galaxies are presented graphically in Fig. \ref{fig:seds_a}
and Fig. \ref{fig:seds_b}.
The data are given in the Appendix as separate tables, together with
the references.   
Although we call them ``SEDs'', we have plotted here the flux (mJy) versus
wavelength, a usual, but conceptually unsound, convention.
We have done so because it is much easier to 
intuitively interpret the trends in the distributions, and 
identify low flux levels because of the vertical scaling
common to all graphs.

We have compared the SEDs shown in Figs. \ref{fig:seds_a} and \ref{fig:seds_b}
with the spectrophotometric models for M\,82 and Arp\,220 given by \citet{silva} 
and \citet{bressan}.
These two galaxies were chosen because they are frequently used as templates
for photometric redshift determinations \citep[e.g.,][]{hughes}.
The models have been normalized by calculating the mean offset over the entire
ensemble of data points in logarithmic flux units.
Such a procedure is equivalent to adjusting the vertical scale so as to
minimize the residuals data$-$model.
The global spectral
normalization is crucial to interpret deviations from the models because 
of the insidious effects of dust extinction, radio free-free absorption, and
different dust temperatures.

Problems with total versus large-aperture flux are apparent
in \ngc\ and He\,2$-$10 (Fig. \ref{fig:seds_a}).
In both cases, excluding the lower-flux points does not change the
normalization relative to the models; we have shown all data
and included them in the best-fit assessment (see below).

\begin{figure*}
\centering\includegraphics[angle=0,width=\linewidth,bb=125 320 480 650]{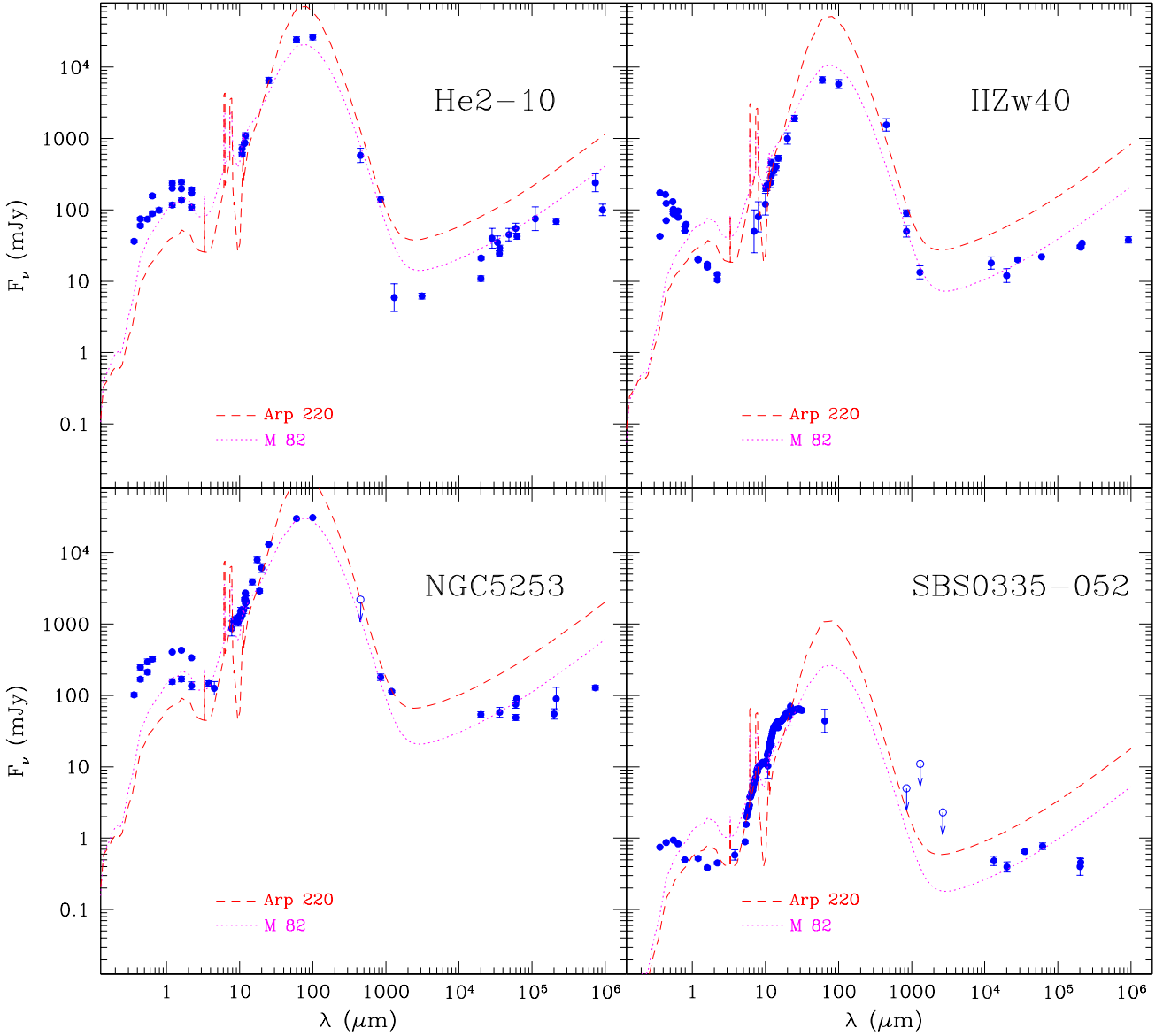}
\caption{Spectral flux distributions for He\,2$-$10, II\,Zw\,40,
\ngc, and \sbs.
Upper limits are shown as open symbols.
Also plotted are the standard templates \citep{silva} for two
prototypical starbursts, M\,82 and Arp\,220.
}
\label{fig:seds_a}
\end{figure*}

\begin{figure*}
\centering\includegraphics[angle=0,width=\linewidth,bb=125 320 480 650]{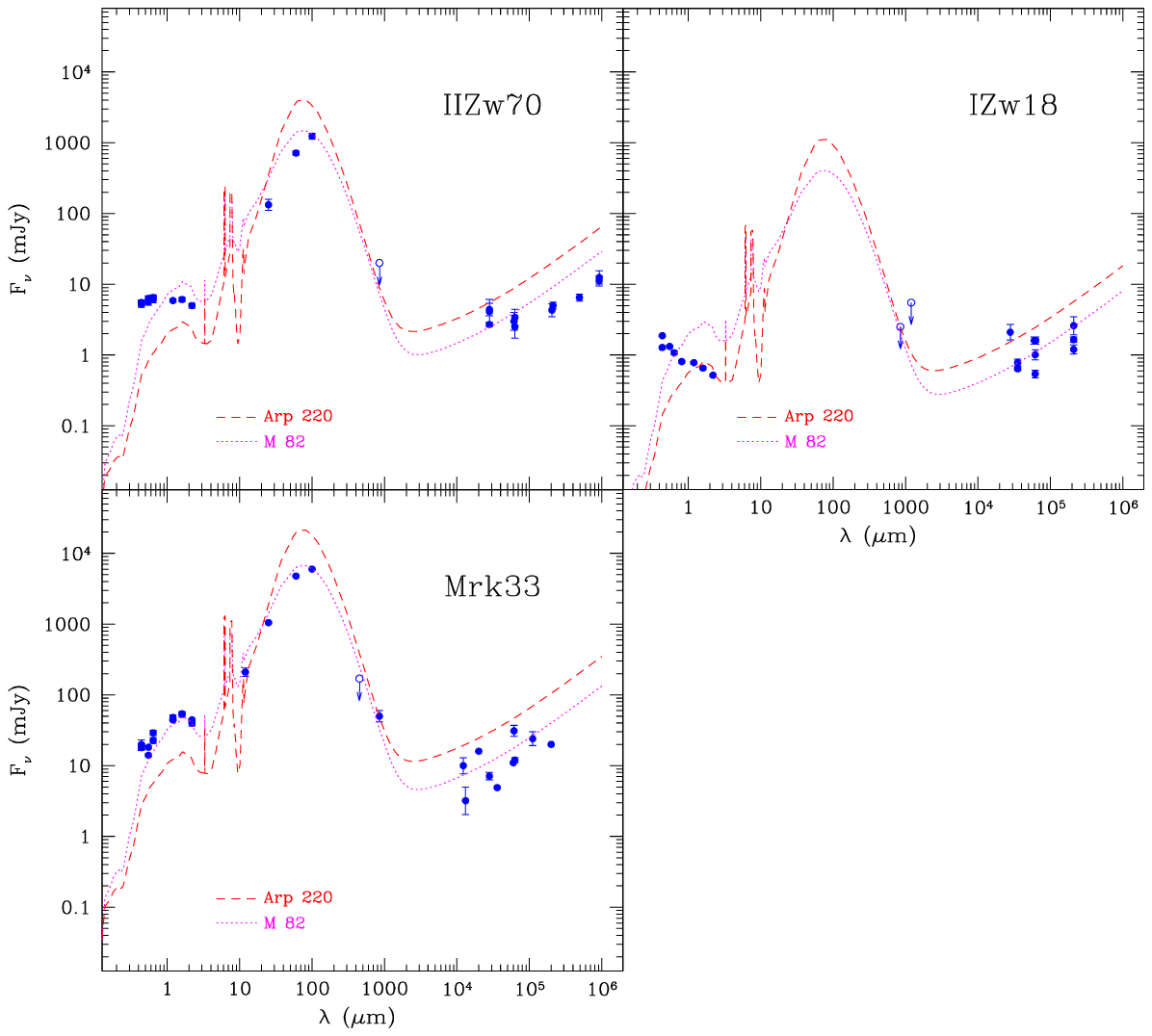}
\caption{Same as in Fig. \ref{fig:seds_a}, but for
II\,Zw\,70, \izw, and Mrk\,33.
}
\label{fig:seds_b}
\end{figure*}

\subsection{Deviations from the standard starburst templates}

To assess how well the SEDs
of M\,82 and Arp\,220 fit the low-metallicity BCDs in our sample,
we calculated the average RMS (logarithmic) residuals after normalization.
We also divided the SED into four spectral regions, and
calculated the average mean offset (normalization) and RMS
residuals individually for each region:
optical$+$NIR ($\lambda\leq 4$\micron),
MIR ($4<\lambda\leq 30$\micron),
FIR-mm ($30<\lambda\leq 7000$\micron), and
radio ($\lambda\geq 7.0$\,mm).
The models were interpolated to the wavelength of the observation.
In every case, the model which gave the smallest RMS residuals after
normalization
is M\,82; Arp\,220 is not a good approximation to any of the galaxies
in our sample.

\begin{figure}
\centering\includegraphics[angle=0,width=\linewidth,bb=27 145 590 650]{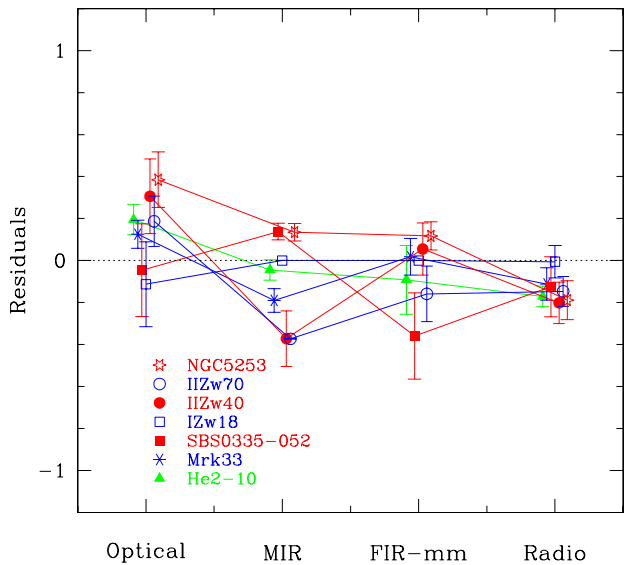}
\caption{The mean logarithmic residuals of each galaxy relative to the best-fit
starburst template, M\,82 as a function of spectral region.
The exact horizontal position is arbitrary, as they have been slightly offset
for better visibility.
Error bars are the error in the mean, so divided by $\sqrt{n}$ relative
to the values in Table \ref{tab:rms}. }
\label{fig:grasilres}
\end{figure}

The results of the normalization/RMS procedure are given in 
Table \ref{tab:rms} for the best-fit model (M\,82).
Column (2) gives the (logarithmic) RMS residuals over the entire SED, and
Columns (3) through (6) contain first the mean offset,
and then the RMS residual for the individual spectral regions.
In all columns, the number enclosed in parentheses is the number
of data points.
The normalization procedure guarantees that
the global mean offset is zero (see previous section).
All units in the table are logarithmic.
Figure \ref{fig:grasilres} shows the deviations from the M\,82 SED
graphically.

\begin{table*}
\caption[]{Deviations from the starburst template, M\,82 \label{tab:rms}}
\begin{tabular}{lcrrrrrrrrrrrr}
\multicolumn{1}{c}{Name} &
\multicolumn{1}{c}{Global RMS} &
\multicolumn{3}{c}{Optical$+$NIR} &
\multicolumn{3}{c}{MIR} &
\multicolumn{3}{c}{FIR-mm} &
\multicolumn{3}{c}{Radio} \\
\multicolumn{1}{c}{(1)} &
\multicolumn{1}{c}{(2)} &
\multicolumn{3}{c}{(3)} &
\multicolumn{3}{c}{(4)} &
\multicolumn{3}{c}{(5)} &
\multicolumn{3}{c}{(6)} \\
\hline
\izw		& 0.44 ~(20) & $-0.11$ &0.57&(8) & \multicolumn{3}{c}{$-$} & \multicolumn{3}{c}{$-$} & $-0.01$&0.24&(10)\\
II\,Zw\,40 	& 0.66 ~(47) & 0.31&0.79&(20)    & $-0.37$&0.48&(13) & 0.06&0.31&(6)    & $-0.20$&0.28&(8)\\
\sbs		& 0.43 (106) & $-0.05$&0.62& (8) & 0.14&0.37&(86)    & $-0.36$&0.36&(3) & $-0.13$&0.35&(6) \\
II\,Zw\,70=UGC\,9560& 0.34 (25) & 0.19&0.38&(10) & $-0.37$& \multicolumn{1}{c}{$-$}& (1) & $-0.16$&0.19&(2) & $-0.15$&0.24&(11) \\
\ngc	        & 0.35 ~(48) & 0.39&0.46&(12) & 0.14&0.20&(23)       & 0.12&0.13&(4)    & $-0.19$&0.26&(8) \\ 
He\,2-10	& 0.29 ~(41) & 0.20&0.29&(16) & $-0.05$&0.12&(6)     & $-0.09$&0.37&(5) & $-0.17$&0.18&(14) \\
Mrk\,33=Haro\,2	& 0.25 ~(29) & 0.13&0.24&(13) & $-0.19$&0.08&(2)     & 0.02&0.15&(3)    & $-0.11$&0.24&(10)\\
\hline
\end{tabular}
\end{table*}

Inspection of Figs. \ref{fig:seds_a}, \ref{fig:seds_b}, and Table \ref{tab:rms}
reveals several important features.
\begin{description}
\item [\it Radio:]
Three of the BCDs (II\,Zw\,40, \ngc, \sbs) have approximately flat global
radio emission; on galaxy scales the radio emission appears to be dominated
by thermal processes.
\item [\it FIR-mm:]
In three objects (II\,Zw\,40, \ngc, He\,2$-$10), the 60--100\micron\ emission
is significantly larger than the best-fit template.
In \sbs, the 65\micron\ point falls well below it.
\item [\it MIR:]
In the objects for which we have considerable spectral coverage
at the mid-infrared wavelengths
(II\,Zw\,40, \ngc, \sbs), the observed MIR spectrum is
not well fit by the M\,82 template.
First, the flux for $\lambda\simgt$12\micron\ {\it exceeds} the template significantly.
Second, there are no PAH features (or Aromatic Features in Emission, AFEs)
in these objects, unlike M\,82 (and Arp\,220). 
\item [\it Optical$+$NIR:]
Three BCDs (II\,Zw\,40, \izw, \sbs) have optical+NIR emission which falls
well below the M\,82 template.
II\,Zw\,70 appears to have an optical excess relative to M\,82,
but a NIR deficit.
\end{description}

\section{Active and passive star formation \label{sec:active}}

The preceding section suggests that
three galaxies deviate from the M\,82 SED over the entire
spectral range covered by our SEDs: 
II\,Zw\,40, \ngc, \sbs.
\izw\ differs significantly in its optical$+$NIR emission, but
the mid- and far-infrared coverage is too sparse to make an
accurate comparison; this situation should be remedied by 
Spitzer.
The remaining galaxies, Mrk\,33 and II\,Zw\,70,
follow the template reasonably well.

The deviant galaxies, II\,Zw\,40, \ngc, and \sbs, share several characteristics, namely:
({\it i})~they host Super Star Clusters (SSCs) \citep{thuan97,calzetti97,beck02};
({\it ii})~unlike most galaxies, their FIR emission peaks at or shortward of 60\micron\ 
\citep{vader,ps};
({\it iii})~there are no PAH features in their MIR spectra \citep{roche,thuan99b,houck04};
({\it iv})~they have $K-L\simgt\,1$ \citep{mg,joylester,hunt01}; 
and ({\it v})~their radio spectra are rising on small spatial scales 
\citep{turner00,beck02,hunt04a}.
The first four characteristics are associated with what has
been called an ``active'' mode of star formation \citep[e.g.,][]{hunt02,hh}.
BCDs which form stars in this way have star-forming regions
which are small (\simlt\,100\,pc) and dense ($n_e$\simgt\,500\,cm$^{-3}$).
Star formation which occurs in larger, more diffuse regions was dubbed
``passive''.
Such a ``dichotomy'' does not occur only in low-mass low-luminosity BCDs;
there is evidence that star formation is ``bimodal'' even in giant 
infrared-luminous galaxies, 
with more intense starbursts tending to be more
heavily obscured and compact \citep{takagi}.

``Active'' regions tend to be more efficient at
shielding ultraviolet (UV) radiation because of the greater concentration of
dust.
Hence the optical depth of the UV photons becomes large, and dust
reaches higher temperatures than it would in more diffuse regions \citep{hhf}.
Because of the high stellar concentration and low metallicities in such BCDs, 
the UV radiation field is also very intense \citep[e.g.,][]{verma}, 
which tends to suppress PAH emission \citep{pl}.
``Passive'' star-forming episodes do not engender SSCs, although they can
generate massive stellar clusters; dust temperatures are lower, and the
MIR spectrum shows conspicuous PAH features. 
The deviant galaxies, II\,Zw\,40, \ngc, \sbs, can all
be classified as ``active'' BCDs;
they are also characterized by compact dense star-forming complexes,
as judged from HST images and optical spectra (Hunt \& Hirashita 2004, 
in preparation).

It is likely that the fifth trait shared by the deviant galaxies 
-compact radio sources with rising spectra-
is also associated with active star formation.
The theoretical basis for the active/passive ``dichotomy'' lies in the size
and density of the star-forming complex \citep{hh}: compact dense regions
form stars ``actively'' and larger more diffuse ones ``passively''.
Rising (inverted) radio spectra surrounding individual stars, 
indicative of dense optically thick \hii\ regions, 
are abundant in the Galaxy \citep{wood},
but in external galaxies have been observed in association
only with compact (\simlt\,10\,pc) massive star clusters 
\citep{kj,beck00,turner00,beck02}.
The inferred densities in these small regions
are extremely high, typically \simgt 2000\,cm$^{-3}$.
Hence, the two properties required for a star-forming complex to be classified
as ``active'' are present in these inverted spectrum sources,
and they may be an integral part of the phenomenon. 

In addition to the three obvious ``deviants'', there may be another
active BCD in our sample: He\,2$-$10.
It contains both SSCs \citep{johnson00}, 
and rising spectrum radio sources \citep{kj}. 
However, its FIR dust temperature is similar to normal spirals,
and its $K-L$ color is $\sim0.8$ \citep{gm}, smaller than
the typical values of ``active'' starbursts \citep{hunt02}.
Its MIR spectrum also contains PAH features, together with 
a deep 10\micron\ absorption trough \citep{roche}.
Hence, He\,2$-$10 may be active on small spatial scales, but 
globally ``passive''. 
At solar metallicity, this galaxy is the most metal-rich 
in our sample.
Its relatively high metal abundance may imply a 
more evolved ISM which
would tend to dilute ``active'' features and
lower the dust temperatures observed over large regions.

It is as yet unclear why some regions form stars actively
and others passively.
Metallicity appears to have little or no influence since
\izw\ forms stars passively \citep{hh}, while \sbs\ forms them 
actively, despite a similar (low) metal abundance.
\citet{hh} speculate that high pressure and shock compression
may be affect the bifurcation into active and passive
regimes.
Star-formation rate (SFR) may also be important, since
active regions tend to have higher SFRs than passive ones; however
this could be an effect rather than a cause.
Active regions are also invariably young (\simlt\,10\,Myr):
indeed optically thick \hii\ regions as signified by warm dust
emission and rising radio spectra appear to be the first
stage of intense star formation 
\citep[e.g.,][]{beck96,kj,vacca02}.
For a region to be defined as ``active'', it may be necessary
to observe it in its early stages.
Passive star formation on the other hand can be either young
or more evolved, and both regimes may be associated with
superbubbles or expanding shells.
Indeed, with the exception of II\,Zw\,70 which is a probable polar-ring
galaxy \citep{cox}, 
all of the galaxies in our sample contain such structures
\citep{martin96,marlowe,martin97,mendez,summers}.
Also with the exception of II\,Zw\,70, 
the sample galaxies host Wolf-Rayet stars \citep{conti91,guseva}, 
implying that the star-formation episodes must be younger than a few 
Myr \citep{schaerer}.

\subsection{Optical/NIR colors }

The hot dust that is associated with active star formation is 
revealed in the $K-L$ color, and is usually
already evident in $H-K$ \citep{hunt02}\footnote{However, if not
accompanied by a bluish $J-H$, then red $H-K$ is probably better ascribed to 
extinction effects.}. 
However, 
optical and NIR colors by themselves may be an ambiguous
diagnostic of the active/passive distinction.
This point can be clarified by examining the colors of our sample galaxies.

\begin{figure}
\centering\includegraphics[angle=0,width=\linewidth,bb=140 160 475 715]{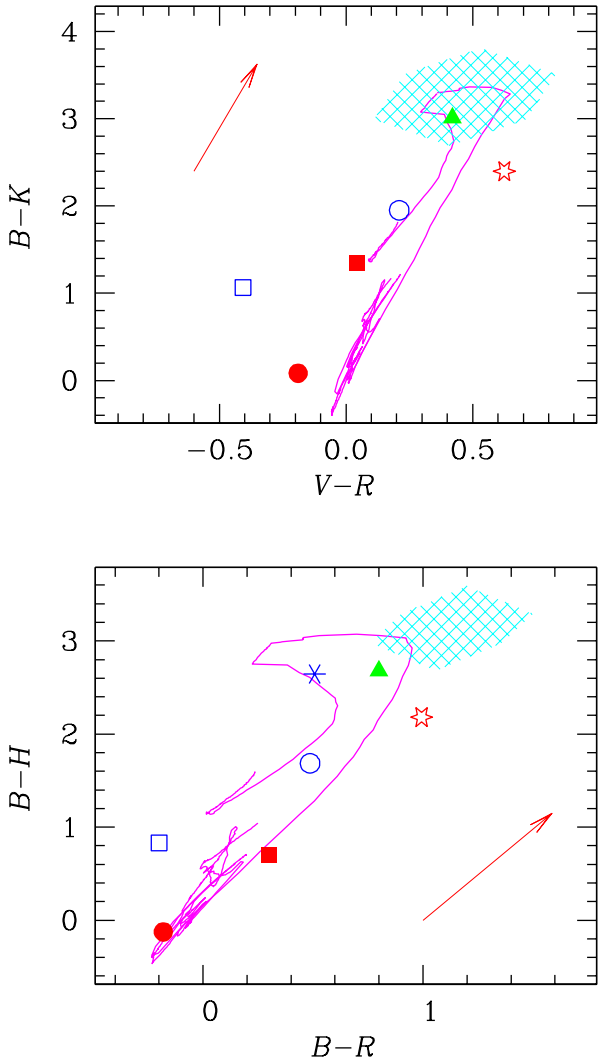}
\caption{Optical and optical/NIR colors of the sample galaxies.
$B-K$ versus $V-R$ are shown in the upper panel, and $B-H$ vs. $B-R$ in the lower.
\izw\ is shown as an open (blue) square;
II\,Zw\,70 as an open (blue) circle;
Mrk\,33 as an (blue) asterix;
He\,2$-$10 as a filled (green) triangle;
\sbs\ as a filled (red) square;
II\,Zw\,40 as a filled (red) circle;
\ngc\ as an open (red) six-point star.
The instantaneous-burst Starburst99 models \citep{sb99} 
for \zsol/20 to \zsol\ are shown as (magenta) solid lines (from 1\,Myr to 1\,Gyr),
and the observed colors of spiral galaxies as a (cyan) grid \citep{dejong96}.
The arrow in both panels shows $A_V\,=$\,1\,mag, according to the
interstellar extinction curve by \citet{cardelli}.
}
\label{fig:colors}
\end{figure}

The BCDs which we have classified as ``active'' (\ngc, II\,Zw,40, \sbs,
and possibly He\,2$-$10) are characterized by a wide range
of optical/NIR colors.
The ``passive'' galaxies on the other hand (Mrk\,33, II\,Zw\,70, \izw)
have uniformly blue to extremely blue optical/NIR colors.
This is shown in Fig. \ref{fig:colors} where
optical/NIR colors are plotted for the sample galaxies.
Also shown in the figure are the Starburst99 models for an instantaneous
burst \citep[][SB99]{sb99}, with 1/20 \zsol\ to \zsol;
the SB99 predictions include nebular continuum emission.
The grid gives the observed range for spiral galaxies \citep{dejong96}.
The \zsol\ curve is the one which touches the spiral galaxy grid at roughly
10\,Myr because of the onset of the asymptotic giant branch.
Fig. \ref{fig:colors} shows that
\izw, \sbs, II\,Zw\,70, and Mrk\,33 are significantly bluer than normal
spirals. 
Only the solar metallicity SB99 models are roughly consistent with
the colors (although none of them predicts the colors of \izw, see \citealt{hunt03}),
and these only for the ages of the models, 1\,Gyr or less.
The remaining galaxies, \ngc, II\,Zw\,40, and He\,2$-$10, are much redder;
they are only slightly bluer (\ngc) or consistent with the colors of early spirals.
Extinction, the effect of which is shown as an arrow in the Fig. \ref{fig:colors}, 
would bring the colors blueward, similar to those of the ``passive'' BCDs.
Therefore, passive BCDs seem to be associated with significantly blue colors
from 0.4 to 2 $\mu$m,
but the colors of active BCDs can vary substantially.
Part of this variation may be due to dust reddening; 
part of it may be an age effect.
With optical/NIR colors alone, this ambiguity is difficult to resolve.

The bluish or normal colors of the active BCDs are in
apparent contradiction with the substantial emission
at $\lambda\,>\,2$\micron\ on a global scale; this is
particularly striking in the case of \sbs, the prototype of an active BCD. 
The seeming paradox can be understood through optical depth 
effects and the morphology at different wavelengths.
The dust extinction is so high on small scales in \sbs\ 
\citep[$A_V\,$\simgt\,12\,mag,][]{thuan99b,ps}
that more than half the
star formation is hidden at 2\micron\ \citep{hunt01}.
Thus, in the optical, we are seeing only the outer shell of the star-forming
regions, the inner parts of which are hidden by dust.
How much of the region is visible depends on the geometry 
of the dust \citep[e.g.,][]{ps}, and on 
the configuration of the emitting regions. 
In the optical, extended ionized gas contributes significantly to the global 
emission in \sbs\ \citep{thuan97},
while in the infrared, the emission is extremely compact, confined to the
most luminous star clusters \citep{dale01b}.

A mixture of these effects is probably responsible for the wide range
of colors in what we have called ``active'' BCDs.
In \ngc, a single cluster
dominates the SED beyond 3\micron, while the entire galaxy contributes
to the optical and NIR emission \citep{vanzi04}.
The same is essentially true for II\,Zw\,40 and He\,2$-$10 in which a large
fraction of mid-infrared and radio emission comes from
compact star clusters in a region less than 200\,pc in diameter 
\citep{kj,beck01,vacca02,beck02}.
Like \sbs, in these objects there is substantial extinction 
\citep[$A_V\,\simgt\,10$\,mag, ][]{kawara,ho90}.
Dust extinction is generally much lower in diffuse extended regions than in 
compact heavily obscured star clusters \citep{calzetti97}.

\section{Modelling the dust in active BCDs \label{sec:dusty}}

The notion that low metallicities imply low dust content in BCDs is
apparently dispelled by our data.
If anything, the active BCDs in our sample have a larger dust
fraction relative to the radio continuum than M\,82.
This is true both for warm dust at MIR wavelengths, and for cooler
dust in the FIR/mm.
Our new SCUBA data allow a more complete estimate of the dust mass
in these BCDs, but first we need to fit the SED.
To do this we have adopted \dusty, a code which solves the radiation transport
problem in a dusty environment \citep{elitzur,dusty}.
The solution is based on a self-consistent equation for the radiative
energy density, and includes the effects of dust scattering, absorption,
and emission.
Such a treatment is important in the active BCDs we study here because
of optical depth effects; simple screen models do not properly account
for dust self-absorption.
Because the code assumes thermal equilibrium, stochastic emission from
very small grains (VSGs) is not considered \citep[e.g.,][]{desert}. 
While this is not a significant problem for dust mass estimates because of the small
contribution of such grains to the total dust mass,
it may affect the parameters we infer from the SED fit
such as temperature and dust distribution.

Because the active BCDs in our sample all host 
compact SSCs or ultradense HII regions or both,
we have assumed a spherical shell geometry for the dust emission.
The inside of
the shell is exposed to radiation from a young stellar population at the
appropriate metallicity and age. 
In general, for any geometry \citep{elitzur},
the scaling properties of the radiative transfer problem 
leads to SED families which, for a given grain type,
are governed by the spatial density distribution of the dust;
the optical depth $\tau$ determines the position within the family. 
As first pointed out by \citet{rr80}, neither the overall luminosity 
nor absolute scales of density or distance are required for the exact solution;
hence, the problem can be defined with relatively few quantities.
With \dusty, it is necessary to specify the temperature \tin\ 
at the innermost boundary of the shell,
and the thickness of the shell in terms of the ratio of the outer to
the inner radius, \yout\ ($\equiv$ \rout/\rin).
For simplicity, we have used one-zone models, so it is necessary to specify
only a single power law index $p$ for the radial run of the mass density,
$R^{-p}$.

\dusty\ currently only supports single-type grains, namely a single size and
chemical composition.
However, it allows the simulation of more realistic grain populations through
the appropriate weighted means.
Thus, to define a grain population, \dusty\ requires the minimum and maximum
grain size, $a_{\rm min}$ and $a_{\rm max}$, assumed to be the same for all chemical
species; the power law index $q$ describing their size distribution, $a^{-q}$;
and their chemical composition.
Several species are supported by \dusty\ including classical silicate
and graphite grains \citep{dl}, and 
amorphous carbons (AMCs) \citep[e.g.,][]{hanner,zubko}.
The fractional number abundances must be specified for each grain type. 

We implemented the SED fitting of our active BCDs by performing a grid search
over various dust density distributions, \yout, dust compositions, and \tin. 
This last imposes the physical size of the innermost shell boundary \rin\ 
because of its dependence
on the intensity of the central source and the normalization of the radial
coordinate.
We defined three basic dust compositions for the grid search: 
({\it i})~a ``standard'' one \citep{dl}; 
({\it ii})~a mixture of silicates, graphite grains, and amorphous carbons
\citep[e.g.,][]{weingartner}; 
({\it iii})~a silicate-free mixture with only graphite grains and amorphous carbons. 
Grain sizes were assumed to be distributed according to a power law, 
with the index $q$ fixed
to $-3.6$, in accordance with \citet{weingartner}, and very similar
to the \citet[][MRN]{mrn} value of $-3.5$.
Grain sizes were taken to be ``standard'' MRN, ranging from
0.005\micron\ to 0.25\micron.
The central exciting source spectrum was taken from SB99
for the appropriate age ($\sim$\,3\,Myr, see Sect. \ref{sec:modelresults})
and metallicity as determined from the optical spectra.
For each metallicity (1/20 and 1/5 \zsol), 
more than 2000 \dusty\ models were generated in the grid,
with power-law indices $p$ ranging from 0 to 1,
\yout\ from 20 to 1000 logarithmically, and \tin\ from 400 to 700\,K.
$\tau_V$ was varied logarithmically from 0.1 to 100.

Because of the importance of thermal radio emission to the observed SED,
we first fit a single power law to
the radio spectrum for $\lambda\geq$ 1\,cm. 
This fit was subtracted from the observed spectrum for $\lambda\,<\,$1\,cm
to obtain the pure dust contribution.
The best-fit \dusty\ model was determined by minimizing the logarithmic residuals
of the radio-subtracted dust spectrum and the \dusty\ models from 5\,$\mu$m to 1\,cm.  
Once the best-fitting \dusty\ model was achieved in this ``coarse'' grid,
the parameter space of the models was refined to describe the SED as well 
as possible.
This resulted in typically another 20$-$50 models being generated.
With the best-fit parameters,
we also investigated the effects of varying the age of the
stellar population and the grain size distribution and chemical composition. 

\subsection{\dusty\ model results \label{sec:modelresults}}

\begin{figure*}
\centering{\includegraphics[angle=0,height=16cm,width=\linewidth,bb=130 355 475 600]{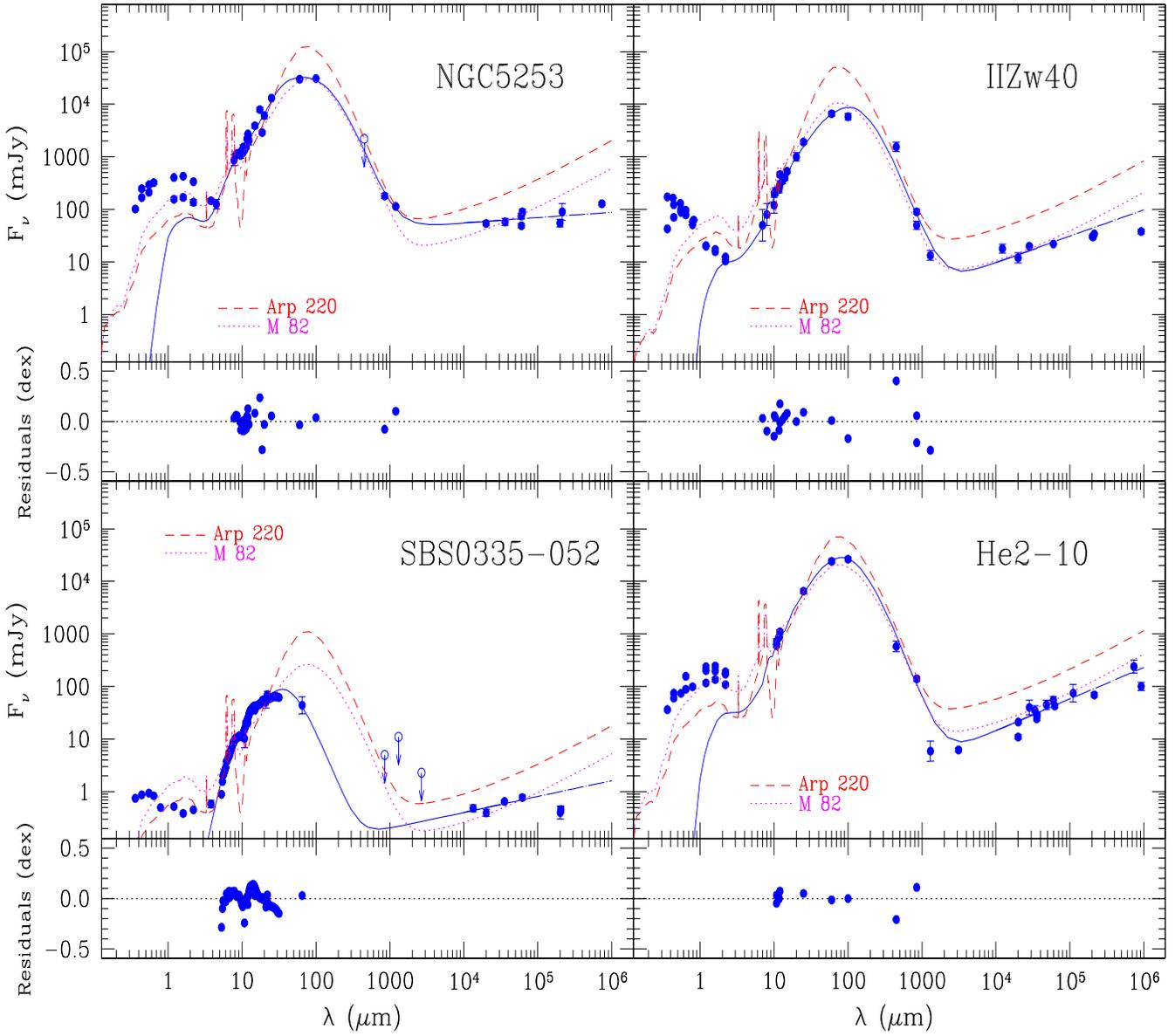}}
\caption{Best-fitting \dusty\ SED models for the ``active'' galaxies (upper panels shown as solid lines),
together with the logarithmic residuals of the fits (lower panels).
The SEDs for Arp\,220 and M\,82 are also shown, repeated from Figures
\ref{fig:seds_a} and \ref{fig:seds_b}.
}
\label{fig:results}
\end{figure*}

\begin{figure}
\centering{\includegraphics[angle=0,width=\linewidth,bb=20 144 590 600]{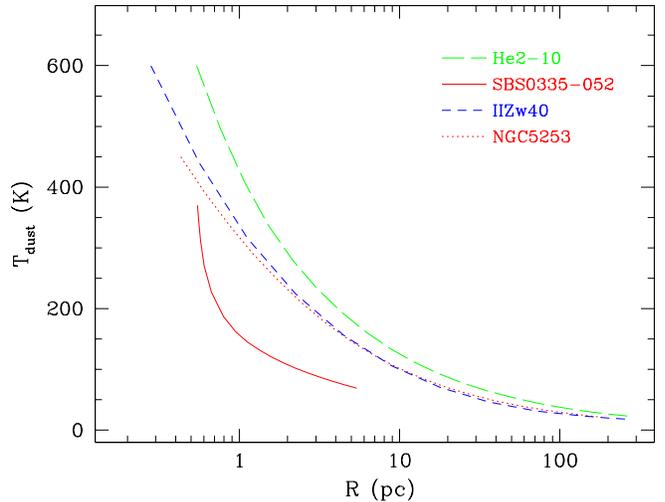}}
\caption{Radial run of dust temperature $T_{dust}$ as predicted
by the best-fit \dusty\ models.
\sbs\ is shown by a solid (red) line,
He2\,$-$10 by a long-dashed (green) one,
II\,Zw\,40 by a short-dashed (blue) one,
and \ngc\ by a dotted (red) one.
}
\label{fig:temp}
\end{figure}

The best-fitting \dusty\ models are shown in Fig. \ref{fig:results},
and the radial run of temperatures in Fig. \ref{fig:temp}.
The dust parameters and radial distributions are reported in Table \ref{tab:dusty}.
Uncertainties in the model parameters are inferred from the grid spacing
with which we searched the parameter space, together with the consequent
variation in the mean logarithmic residuals.
The uncertainties in the temperature at the inner shell boundary \tin\ is
roughly 25$^\circ$\,K, and that at the outer boundary \tout\ on the order
of a few (2-5$^\circ$\,K).
As for the dust mass distribution,
the power-law index has an uncertainty of a few tenths, 
and the absolute values of the radii between 10 and 20\%.
The dust optical depth $\tau_V$ is precise to $\sim$30\% for the larger
values ($\tau_V\,=\,30$) and
about 20\% for the smaller ones ($\tau_V\,\simlt\,20$).

The optical and NIR data points were not included in the fits because of
unknown extinction and geometry effects.
The stellar populations reflected by these data in the global SEDs
are almost certainly not subject to the high extinction experienced
by the central star clusters responsible for heating the
dust \citep[e.g.,][]{calzetti97}.
Indeed, Fig. \ref{fig:results} shows that if they were, they would not be
visible at all at optical wavelengths. 

Also given in Table \ref{tab:dusty} are the
inferred dust masses \md\ and infrared luminosity \lir\
from 5\,$\mu$m to 1\,cm; this last is approximately the bolometric luminosity
because of dust reprocessing of the incident UV radiation.
\lir\ is easily obtained by integrating the model over the fitted wavelength
range, and applying the normalization factor used to rescale the \dusty\
models to the observed SEDs.
The dust mass is less straightforward to derive from the models.
Although \md\ is proportional to the optical
depth $\tau_V$ \citep[see e.g.,][]{ps}, 
the former depends on the {\it amount} of dust and its
geometry, while
the latter is also a function of dust composition, its optical properties, and 
the grain size distribution.
Hence, converting the optical depths given by the models into dust mass
involves calculating weighted averages of these quantities in the same way 
as in the \dusty\ code.
We have done this as well as possible, but the approximations
for the optical quantity weighting could introduce an additional 10-20\% 
uncertainty in the dust masses.
The equations which relate $\tau_V$ to \md\ are given in
Appendix \ref{app:equations}.
The main uncertainty in \md\ however is the dust geometry,
that is to say the radial distribution and truncation radius \rout.
From the difference between the best fit and the next best one,
uncertainties in the dust mass are usually 20-30\%, but can be as high
(in the case of \sbs) as a factor of 3.

\begin{table*}
\caption[]{\dusty\ best-fit parameters for ``active'' BCDs\label{tab:dusty}}
\begin{tabular}{lrcrrrrrcrrr}
\multicolumn{1}{c}{Name} &
\multicolumn{1}{c}{$\alpha$} &
\multicolumn{1}{c}{$R^p$} &
\multicolumn{1}{c}{\tin} &
\multicolumn{1}{c}{\tout} &
\multicolumn{1}{c}{\rin} &
\multicolumn{1}{c}{\rout} &
\multicolumn{1}{c}{$\tau_V$} &
\multicolumn{1}{c}{Dust$^a$} &
\multicolumn{1}{c}{\lir} &
\multicolumn{1}{c}{\md} &
\multicolumn{1}{c}{$\langle \sigma \rangle^b$ ($N$)} \\
&(radio)&&\multicolumn{1}{c}{(K)} & \multicolumn{1}{c}{(K)} &
\multicolumn{1}{c}{(pc)} & \multicolumn{1}{c}{(pc)} &&&
\multicolumn{1}{c}{(\lsol)} & \multicolumn{1}{c}{(\msol)} & \multicolumn{1}{c}{(log)}\\
\hline
II\,Zw\,40      &-0.5 & $R^{-0.25}$ & 600 & 18 & 0.3 & 297 & 20 & WD & $3.0\times10^9$ & $1.8\times10^5$ & 0.15 (19)\\
\sbs		&-0.3 & $R^{-1}$   & 400 & 69 & 0.5 & 5.4 & 30 & WD$+$Zb & $1.4\times10^9$ & $1.1\times10^2$ & 0.08 (88)\\
\ngc	        &-0.2 & $R^{-0.3}$ & 450 & 19 & 0.4 & 213 & 10 & Grf$+$Zb & $1.8\times10^9$ & $1.6\times10^5$ & 0.10 (25)\\
He\,2-10	&-0.6 & $R^0$      & 600 & 23 & 0.4 & 200 & 20 & WD      &$5.6\times10^9$ & $9.6\times10^4$ & 0.09 (10)\\
\hline
\end{tabular}

\smallskip
$^a$~Dust is coded by: WD$+$Hn (0.3 silicates, 0.4 graphite, 0.3 Hanner AMCs);
WD$+$Zb (0.3 silicates, 0.4 graphite, 0.3 Zubko et al. AMCs);
Grf$+$Zb (0.70 graphite, 0.3 Zubko et al. AMCs);
WD (0.6 silicates, 0.4 graphite).
All grains follow a power-law size distribution with $q=3.6$ and $a_{\rm min}=0.005\,\mu$m,
$a_{\rm max}=0.25\,\mu$m, except for Grf$+$Zb which has $a_{\rm min}=0.03$ and $a_{\rm max}=0.3\,\mu$m. \\
$^b$~Mean residuals from 5\,$\mu$m to 1\,cm.
\end{table*}

The most important result of the fitting procedure is the good quality
of the fits:
with the exception of II\,Zw\,40, the mean (logarithmic) residuals
are 0.1 or better (see Table \ref{tab:dusty}).
This implies that both the assumed shell geometry and the approximation of
thermal equilibrium for the dust emission may not be unrealistic hypotheses.
On the other hand, there could be a strong degeneracy
between the temperature distribution of the dust derived from the
assumption of thermal emission and the stochastic emission from VSGs.
Either way the dust masses are likely to be unaffected because of the
neglible contribution from the small-grain population.  
In II\,Zw\,40, the quality of the fit (mean residuals 0.15) is almost certainly 
affected by aperture effects since the cool-dust emission at 850\,$\mu$m appears 
extended beyond 40\arcsec.
The 230\,GHz point of 5.9\,mJy \citep{kobulnicky} for He\,2$-$10 is impossible
to reconcile with any fit and was disregarded both in the normalization 
and in the calculation of the residuals.

In general,
with the exception of \sbs, we find that
the dust mass distribution is not strongly peaked;
either uniform or low-index power laws provide the best fit for the radial 
distribution of the dust.
Again with the exception of \sbs, 
the star clusters modelled by our fits 
have rather thick shells with \rout/\rin$\sim500-1000$. 
They are also rather extended,
having radii of $\simgt$200\,pc \citep[see also][]{vanzi04}.
However, this size is deceptive, being the truncation radius of
the dust shell; the dust at these radii is cool,
around 20$^\circ$\,K (see Table \ref{tab:dusty} and Fig. \ref{fig:temp}), and would
not be detected in the MIR.
The form of the SEDs at longer IR wavelengths dictates the presence of such
cool dust, because of their broad peak at $\sim$60\,$\mu$m.
On the other hand, as shown by the IRS spectrum of \citet{houck04},
the dust in \sbs\ is very warm, peaking at $\sim$30\,$\mu$m.
It is also very compact; our fit gives a diameter of $\sim$\,10\,pc,
with a dust temperature of $\sim70^\circ$\,K at this boundary.
Our new 65\,$\mu$m ISOPHOT flux of 44\,mJy is consistent with this,
but combined with the Spitzer IRS spectrum
results in a much different \dusty\ fit than obtained previously
\citep[][see below]{ps}.

The dust in these BCDs also appears to be relatively ``standard''.
The same ($\sim$MRN) grain size distribution  and standard composition
(0.3 silicates, 0.4 graphite grains, 0.3 AMCs), 
fits all galaxies except for
\ngc, in which there seems to be a paucity of small grains and silicates,
in agreement with \citet{vanzi04}.
AMC grains are necessary in \sbs\ and \ngc,
but not in He\,2$-$10 and II\,Zw\,40 where a
``standard'' MRN composition provides the best fit. 

We found very little if any dependence of the models on either the
metallicity or the age of the exciting stellar cluster.
This is probably because luminosity enters
into the radiative transfer problem only through
the flux at the inner radius of the shell; once this
is specified by the temperature \tin\ and the radius \rin,
the problem merely is rescaled through normalization.
In all cases, we therefore adopted a 3\,Myr cluster for the final
best fit, consistent with the optical spectra and the presence of Wolf-Rayet
features (see Sect. \ref{sec:active}), although we used the SB99 stellar population with
the metallicity closest to that of the particular BCD.

\subsubsection{II\,Zw\,40}

Our model gives a total IR luminosity \lir\ of $3.0\times10^9\,$\lsol, in excellent
agreement with the OB stellar luminosity $L_{\rm OB}$ of $3\times10^9\,$\lsol\
reported by \citet{beck02}, and with earlier estimates 
\citep{wynnwilliams}.
The structure in the starburst of II\,Zw\,40 is however not well approximated by a single
dusty shell, since high-resolution radio maps show four compact nebulae
separated by 10-12\,pc \citep{beck02}.
The inner radius of 0.3\,pc given by our best-fit model is probably the
best guess for the combined emission from these clusters;
there is clearly warm dust (our model gives \tin\,=\,600\,$\pm\,$25\,K) heated by
the massive stars.

Assuming an age of 3-5\,Myr, we can use the SB99 models to derive the stellar mass
responsible for the IR luminosity 
$3.0\times10^9\,$\lsol\ gives a total stellar mass of $2.0\times10^6\,$\msol\
associated with the starburst.
This is again in very good agreement with the estimate of \citet{beck02} who find
a total of $2\times10^6\,$\msol\ in young stars, and is $\sim$\,10\% of the 
total stellar mass estimated by \citet{vanzi96}.

The cool-dust emission is clearly extended in this galaxy, since the flux
at 40\arcsec\ is about half of the global emission in the SCUBA image.
Narrow-band images and long-slit spectroscopy show that the emission-line
region in II\,Zw\,40 is more extended than 40\arcsec\ \citep{vanzi96}, so the 
presence of spatially extended cool dust in the SCUBA image is not surprising.
The diameter of $\sim$600\,pc (where $T\,\sim\,20$\,K)
given by our model is much smaller than this $\sim10$\arcsec; 
this should be however roughly the maximum extent of the dense
star clusters that comprise the starburst, the core alone of which has an extent of 
$\sim$150\,pc \citep{beck02}. 

The extinction given by our model of $\tau_V\,=\,20\,$mag is significantly higher
than previous estimates \citep[$\tau_V\sim3-5$; ][]{joylester,walshroy,vanzi96}.
Part of the reason for this may be the assumption of a foreground screen by
these authors. However, the discrepancy may also be ascribed to substantial
optical thickness; in very dusty star clusters, even the NIR hydrogen recombination
lines do not probe the central most obscured regions \citep[e.g., \sbs;][]{hunt01}.

The best-fit radio index is $\alpha=-0.2$ which gives large \dusty\
residuals $\langle \sigma \rangle\,=\,0.2$.
We therefore steepened the
radio spectral index $\alpha$ gradually to  $-0.5$; 
the residuals monotonically decrease to 0.15 as $\alpha$ steepens.
We therefore adopted $\alpha=-0.5$ which
clearly does not well approximate the radio emission at lower frequencies;
there is evidently substantial free-free absorption in II\,Zw\,40
\citep[see][]{beck02}. 

\subsubsection{\sbs}

Our \dusty\ fit for \sbs\ is substantially different than that presented by
\citet{ps}; most importantly, it predicts a much lower dust mass and a much smaller shell.
Our dust mass estimate of $1.1\times10^2$\,\msol\ is a factor of 10 lower than
by \citet{houck04}, and $10^3$ times lower than previous
estimates \citep{thuan99b,hunt01,ps}.
There is roughly a factor of 3 uncertainty on \md; our next best fit gives
a dust mass of $\sim3.7\times10^2$\,\msol\ with a truncation radius \rout\ of 11\,pc.
However,
if \rout\ (hence \md) are much larger, the residuals become unacceptably high.
The main reason for the difference
is the input SED in which we have included the Spitzer IRS
spectrum given by \citet{houck04}, and our new ISOPHOT 65\,$\mu$m photometry
(44\,mJy vs. 112\,mJy).
As pointed out by \citet{houck04}, the spectral turnover at $\sim30$\,$\mu$m
in the IRS data implies that the cool dust is warmer than previously thought,
lowering the inferred dust mass. 
The IR luminosity $1.4\times10^9$\,\lsol\ of our model is also $\sim$3 times lower than that of
\citet{ps}, again because of the different SED.
The definitive answer of how much dust is contained in the SSCs of \sbs\  
must wait for the long-wavelength Spitzer MIPS photometry.

The other parameters of our best-fit \dusty\ model are also rather different
from those by \citet{ps}.
Although the radial fall-off of their three-zone model is the same as our one-zone
($R^{-1}$) and the optical depths are the same ($\tau_V\,=\,30$),
they find a much thicker and hotter shell than we do (\tin\,=\,700\,K, \yout\,=\,1000).
Both of these differences can also be ascribed to the new IRS data.
The smaller thickness of our model is due to the lack of cool dust emission beyond
30\,$\mu$m; thick shells emit at longer wavelengths because of the greater abundance
of cooler dust.
The lower temperature of 400\,K is dictated by the slope and relative flux of the 
IRS spectrum for $\lambda\simlt$\,10$\mu$m;
hotter dust (than $\sim$500\,K) gives too much flux, 
and a slope that is not sufficiently steep.
The dust parameters we find for \sbs\ also differ from those found by \citet{ps}.
Although the dust composition of both fits is similar,
they find evidence for a flatter exponent in the size distribution and larger
grains, while we find that it is possible to fit the SED with ``standard'' dust
sizes.

Assuming an age of 3-5\,Myr, as for II\,Zw\,40,
we can derive the total stellar mass responsible for the IR luminosity 
$2.8\times10^9\,$\lsol\ and find $\sim10^6\,$\msol,
only slightly lower than that found by \citet{ps}, $2\times10^6$\,\msol.
It is however
much lower than the $6.6\times10^6$\,\msol\ estimate by \citet{hunt01}
on the basis of \bra. 
Radio measurements have shown that the number of ionizing photons inferred from \bra\ 
in \sbs\
is inconsistently high \citep{hunt04a} because of the influence of stellar winds; 
the stellar mass given by \citet{hunt01} is consequently overestimated.
Indeed, embedded massive star clusters and SSCs all tend to have masses around
$\sim\,10^6$\,\msol \citep{gallagher,smith01,vacca02,degrijs}, and this value is typical
of the BCDs in this paper.

\subsubsection{He\,2$-$10}

The total \lir\ of our best-fit model is $5.6\times10^9$\,\lsol,
in good agreement with the value of $6\times10^9$\,\lsol\ given
by \citet{vacca02}.
Our dust mass estimate \md\,$\sim\,10^5$\,\msol\ is substantially
lower than theirs, although consistent with that of \citet{baas}.
The stellar mass inferred from SB99 assuming an age of 3\,Myr is
$4\times10^6$\,\msol,
consistent with previous estimates \citep[e.g.,][]{johnson00,vacca02}

\citet{vacca02} have modelled the IR emission of He\,2$-$10 
using optically-thin shell models of the dust with a uniform density distribution.
Despite the differences in the model assumptions,
\rout\ of our \dusty\ model is similar to theirs (100-200\,pc for the
various emission knots).
However, the inner and outer temperatures are rather different:
we find \tin\,=\,400\,K and \tout\,=\,23\,K, while they find a factor of two cooler dust
at the inner shell boundary and warmer dust ($\sim30$\,K) at the outer
boundary.
They also find much lower extinction ($A_V\simlt8$) while we find
$\tau_V\,=\,20$, a value consistent with the estimate by \citet{phillips} and \citet{kobulnicky95}.
The differences in our results and those of \citet{vacca02} are almost certainly due to
their assumption of an optically thin shell, and a single grain radius of 0.1\,$\mu$m.
The high value of $\tau_V$ given by our model is consistent with the
observation of \citet{vacca02} that ``none of the SSCs seen in the HST UV
and optical images, or the ground-based NIR images, is detected in the
Gemini $N$-band image, and vice versa''.

\subsubsection{\ngc}

The IR luminosity of $1.8\times10^9$\,\lsol\ given by our \dusty\ model is in
excellent agreement with that of $1.7\times10^9$\,\lsol\ of \citet{gorjian01},
and of \citet{vanzi04} after correcting for the different adopted distance.
This translates into a stellar mass of $1.2\times10^6$\,\msol, assuming an age
of 3-5\,Myr and using the SB99 models as before.
As mentioned previously, this age is appropriate for galaxies such as \ngc\ with
Wolf-Rayet features in their optical spectrum \citep{schaerer97}.
Our estimate of the stellar mass agrees well with that of \citet{calzetti97},
and again is typical of the mass of embedded clusters and SSCs (see above).

The SED of \ngc\ has also been modelled with \dusty\ by \citet{vanzi04}.
The dust mass \md\ of $1.6\times10^5$\,\msol\ given by our best fit is
consistent with, although lower than, theirs ($2.3\times10^5$\,\msol\ with
our Cepheid distance of 4.1\,Mpc).
The geometry of the two fits is also similar, with
our one-zone best-fit model $R^{-0.3}$ compared to
their two-zone distribution, $R^0$, $R^{-0.5}$.
The outer shell radii also agree very well, once the different adopted
distance has been accounted for (216\,pc vs. our value of 213\,pc).
Moreover, the two dust compositions are similar, with a lack of silicates,
and 70\% (us) to 80\% (them) in graphite grains.
The optical depth $\tau_V\,=\,10$ of our best-fit model is in good agreement 
with that of $>9$\,
mag measured by \citet{calzetti97} from optical hydrogen recombination lines,
although slightly higher than ($A_V\,=\,8$) derived by \citet{vanzi04}.

However, \citet{vanzi04} find substantially hotter dust (570\,K vs. 450\,K)
and slightly larger grain sizes.
The differences between our results and those of \citet{vanzi04}
are almost certainly attributable to the differences in fitted grain properties. 
Our second-best fit (also characterized by 
$\langle \sigma \rangle\,=\,0.10$) is obtained
with a ``standard'' grain size distribution (see Sect. \ref{sec:dusty} and
Table \ref{tab:dusty}), and
results in a dust mass \md\ of $4.8\times10^4$\,\msol, with $\tau_V\,=\,8$
and \tin\,=\,400\,K.
This dust mass is more than three times smaller than that obtained with
our best-fit model which has $a_{\rm min}\,=\,0.03\mu$m (rather
than the ``standard'' $a_{\rm min}\,=\,0.005\mu$m).
Indeed,
the mass of dust with given flux from thermalized grains depends
linearly on the grain size \citep[see e.g.,][]{hildebrand}.

The uncertainties in the dust parameters used for the \dusty\ fits are probably
the largest contribution to the uncertainties in the derived parameters such
as \tin\ and \md.
With the models and the data we have, it is difficult to 
determine grain size distributions.
The same SED can be equally well fit by very different grain populations.
UV observations and more detailed modelling of dust emission are both
needed to better constrain grain size distributions and dust chemical
composition.

\section{Scaling laws revisited \label{sec:scaling}}

As mentioned in the Introduction,
there is previous evidence that neither IR nor radio luminosity accurately
traces the SFR in the low-luminosity low-metallicity galaxies
studied here.
In this section, we examine how well the canonical correlations
among MIR, far-IR, and radio luminosities are obeyed by the BCDs in our sample.

\subsection{Radio/FIR correlations \label{sec:radiofir}}

Several previous authors have noted that the ubiquitous radio to IR correlation
\citep[e.g.,][]{condon92} does not hold for some low-metallicity
dwarf galaxies \citep{dale01b,beck02}
or galaxies with young starbursts \citep{roussel03}.
Such objects are characterized by high IR--radio ratios, and the
common explanation is a lack of the synchrotron emission 
which typically dominates the radio continuum of luminous large spirals
\citep[e.g.,][]{price,roussel03} or free-free absorption at 1.49\,GHz \citep{condon91}.
Both of these effects are at work in the BCDs in our sample, judging
from our fitted radio indices (see Table \ref{tab:dusty})
and previous detailed radio observations \citep{kj,turner00,beck02,hunt04a}.

First, we examine the ``$q$ ratio'' of the galaxies for which
we have adequate data.
$q$, defined by the logarithm of the ratio of the far-IR flux and the radio flux at 1.4\,GHz
\citep{helou85,condon92}, for most galaxy populations is
very robust: $\langle q \rangle\,=\,2.34\pm0.19$ \citep{condon91}.
He\,2$-$10 and \ngc\ have the highest $q$ values in our sample with $q\,\approx\,2.6$
\citep[see also ][]{vacca02,roussel03}.
These values are typical of ultraluminous infrared galaxies, 
thought to be powered by dense compact starbursts
which are optically thick at 1.49\,GHz \citep{condon91}.
The $q$ values for Mrk\,33, II\,Zw\,40, and II\,Zw\,70 (2.5, 2.4, 2.3, respectively)
are apparently representative of the normal galaxy population.
\sbs\ has a rather low $q$, 2.0, which is due to the reduced IR emission
longward of 30\,$\mu$m.
The most ``deviant'' galaxies (He\,2$-$10 and \ngc\ on the high side,
and \sbs\ on the low side) differ \simgt1$\sigma$ from the mean value;
thus our results neither confirm nor rule out the applicability to BCDs of
the total FIR$-$1.4\,GHz correlations.
Apparently BCDs are not extreme cases {\it globally} \citep{roussel03},
although radio and IR emission do show strong deviations from the correlation
on small spatial scales \citep[e.g.,][]{turner00,beck02}.

Next, we examine the ratios of the 60\,$\mu$m and radio luminosities
that are necessary for the applicability of the \citet{condon92}
scaling relations for SFR. 
In order for the 60\,$\mu$m--1.4\,GHz luminosities
to give the same SFR, the ``standard'' SED must have
$f_{60\mu {\rm m}}/f_{1.4{\rm GHz}}\,=\,111$ \citep{hopkins02}.
With the exception of \sbs\ and II\,Zw\,70, 
all galaxies with available data exceed this value by factors of 2 to 3.
The SFRs inferred from the 60\,$\mu$m luminosity are thus 2 to 3 times higher
than those from the radio, and several times higher (1-3\,\msol\,yr$^{-1}$) than
the median value of 0.3 \msol\,yr$^{-1}$ found by \citet{hopkins02} for 
a similar sample of BCDs.
If we assume that all the radio emission is thermal, rather than the 10\%/90\%
thermal/non-thermal mix at 1.4\,GHz \citep{condon92}, 
we would infer even higher SFRs: 2--14\,\msol\,yr$^{-1}$. 
Table \ref{tab:sfr} gives the SFRs inferred from the IR and radio
luminosities under standard assumptions (see also Sect. \ref{sec:firmir}).
The thermal radio estimate for SFR is only valid for those BCDs such as \ngc, II\,Zw\,40,
and \sbs\  with rather flat radio spectra.
The deviations from the ``standard'' radio--IR correlation appear to be
greater for the ``active'' BCDs, although our sample is small, and
we have few passive galaxies with which to compare.

These results are in contrast with those of \citet{hopkins02} who
found that SFRs inferred from 60\,$\mu$m and 1.4\,GHz luminosities are
in close agreement for local BCDs.
The main reason for the difference may be due to the paucity
of data in their sample: of 50 sources, only 18 were detected with IRAS at
60\,$\mu$m, and 23 at 1.4\,GHz.
Both the radio and IR data are flux-limited, with limits of 1--2\,mJy for the radio
surveys (NVSS and FIRST) and 200\,mJy for IRAS \citep{hopkins02}.
The ratio of these limits is very close to the prediction of the
IR--radio correlation, perhaps making it possible to mimic the common trend.
Moreover, few of the lowest-metallicity galaxies are detected at both IR and radio
wavelengths. 
Of the $\sim$20 BCDs in their sample with $Z<$\zsol/10, only three have both detections;
two of these three have IR-inferred SFRs greater than radio ones.

The extremely metal poor ``active'' BCD \sbs\ apparently does follow the IR--radio 
correlation; the SFRs inferred from the 60\,$\mu$m and 1.4\,GHz luminosities
are the same, 0.2\,\msol\,yr$^{-1}$. 
However, the true SFR is $\sim$\,1\,\msol\,yr$^{-1}$ \citep{hunt04a}.
The 60\,$\mu$m luminosity is anomalously low because of the absence of cool 
dust, and the radio emission is low because of the strong thermal contribution 
and free-free absorption at 1.4\,GHz. 
This is yet another aspect of the ``conspiracy'' described by \citet{bell03}:
in low-luminosity dwarfs,
``the radio--IR correlation is linear not because both radio and IR emissions
track SFR, but rather because both radio and IR emissions fail to track SFR
in independent, but coincidentally quite similar, ways'' \citep[see also][]{price}. 

Starburst age can be another factor in determining the validity of the IR--radio 
correlations, and the ability of these luminosities to accurately trace SFR.
In their studies of star-forming galaxies,
\citet{roussel03} and \citet{cannon04} 
have found that the radio continuum emission in young ($<$10\,Myr) starbursts
tends to be pure thermal.
Indeed, \citet{cannon04} have suggested that the observed radio spectral
index be used as an age indicator.
Heavily embedded star clusters will have positive spectral indices
at ages $\simlt$\,1\,Myr, associated with
optically thick thermal bremsstrahlung \citep[e.g.,][]{beck00}. 
Then as they become increasingly optically thin, the spectral index will
flatten to $-0.1$.
When supernovae (SN) and supernova remnants (SNRs) begin to contribute to
the radio emission after $\sim$10\,Myr, the radio index will steepen.
Evolved starbursts in
spiral galaxies in which cosmic-ray diffusion processes have taken over
($\simgt$50--100\,Myr) will have the ``canonical'' radio spectral index
of $\sim-0.8$ \citep{condon92}.

\subsection{MIR/FIR correlations \label{sec:firmir}}

The availability of ISO data for large numbers of galaxies
has prompted the use of MIR luminosities to trace star formation
\citep[e.g.,][]{roussel01,forster04}.
Moreover, the relative constancy of the 7 to 15\,$\mu$m
ratio in normal star-forming galaxies has led to the
formulation of ``standard'' SEDs, and the potential application
to high-redshift sources \citep{dale01a,dale02}.
The good correlation between PAH strength and cold dust in
normal star-forming disks \citep{haas} is one of the tenets of 
such formulations, but it is not clear that PAH strength
can be used to unambiguously trace star formation.
In Sect. \ref{sec:seds}, we found that the ``active'' BCDs
were not well fit by the starburst template, M\,82.
Here we examine the implications in the context of
MIR-diagnostics of SFRs and MIR flux ratios.

Because of the absence of PAH features or AFEs in the active BCDs,
and the presence of warm dust,
such galaxies do not follow the usual trends for MIR emission.
In \ngc, II\,Zw\,40, and \sbs, 
the 15 to 7\,$\mu$m flux ratio is $\simgt$3 times larger
than would be expected from
the correlations presented in \citet{roussel01}.
These authors emphasize that their MIR correlations are only
applicable in star-forming regions which are dominated by
AFEs, but without spectral information
it is difficult to assess this a priori. 
Moreover, 
the ratio of 15 to 850\,$\mu$m flux in \ngc, II\,Zw\,40,
and He\,2$-$10 exceeds that predicted by the \citet{dale02}
models by a factor of $\sim3$;
in \sbs, where we have no SCUBA detection, even the lower limit
is greater than the ``standard'' prediction. 
We have no data for the 7\,$\mu$m flux of He\,2$-$10, but the
remaining active BCDs show
a similar trend for the 7 to 850\,$\mu$m flux ratio.

Again because of the suppression of AFEs in the active BCDs,
the SFRs predicted by their 7\,$\mu$m luminosity \citep{roussel01}
are factors of $\simgt9$ lower than those predicted by the (thermal)
radio emission.
In II\,Zw\,40, the deficit is $\simgt$50.
Their 7\,$\mu$m SFRs also do not compare well with those inferred from
the 15\,$\mu$m luminosities, and can be as much as a factor of $\simgt$10 lower. 
Table \ref{tab:sfr} gives the SFRs inferred from the MIR, FIR, and radio
luminosities, and enables a comparison of the different indicators which
hold for other samples. 
It is clear that the active BCDs in our sample are highly discrepant,
relative to normal star-forming galaxies.
On the other hand, certainly II\,Zw\,70 and probably Mrk\,33
show SFRs that scale as usual.

\begin{table*}
\caption[]{SFRs$^a$ from various luminosity indicators \label{tab:sfr}}
\begin{tabular}{lccccc}
\multicolumn{1}{c}{Name} &
\multicolumn{1}{c}{SFR$_{7\mu{\rm m}}^b$} &
\multicolumn{1}{c}{SFR$_{15\mu{\rm m}}^b$} &
\multicolumn{1}{c}{SFR$_{60\mu{\rm m}}^c$} &
\multicolumn{1}{c}{SFR$_{{\rm 1.4GHz}}^d$} &
\multicolumn{1}{c}{SFR$_{{\rm 1.4GHz}}^e$} \\
\multicolumn{1}{c}{(1)} &
\multicolumn{1}{c}{(2)} &
\multicolumn{1}{c}{(3)} &
\multicolumn{1}{c}{(4)} &
\multicolumn{1}{c}{(5)} &
\multicolumn{1}{c}{(6)} \\
\hline
\izw		& \multicolumn{1}{c}{--} &\multicolumn{1}{c}{--} &\multicolumn{1}{c}{--} & 0.04 & 0.35\\
II\,Zw\,40 	& 0.1 & 1.1 & 1.2 & 0.6 & 5.4 \\
\sbs		& 0.2 & 1.8 & 0.2 & 0.2 & 1.8 \\
II\,Zw\,70=UGC\,9560 &  \multicolumn{1}{c}{--} & \multicolumn{1}{c}{--} & 0.4 & 0.3 & 2.4 \\
\ngc	        & 0.1 & 0.9 & 0.7 & 0.2 & 1.9 \\
Mrk\,33=Haro\,2	& \multicolumn{1}{c}{--} & \multicolumn{1}{c}{--} & 3.4 & 1.6 & 13.5 \\
He\,2-10	& \multicolumn{1}{c}{--} & 2.6 & 2.5 & 0.8 & 6.9 \\
\hline
\end{tabular}

\smallskip
$^a$\msol\,yr$^{-1}$. \\
$^b$Assuming the relations in \citet{roussel01}.\\
$^c$Total SFR, assuming the relation in \citet{hopkins02}.\\
$^d$Total SFR, assuming the relations in \citet{condon92}.\\
$^e$Assuming the radio emission is purely thermal free-free at 10000\,K \citep{condon92}.
\end{table*}

Another assessment of the validity of ``standard'' SEDs to BCDs 
is the 24\,$\mu$m $q$ ratio \citep{appleton}.
The 24\,$\mu$m-to-radio correlations in the Spitzer Extragalactic First Look Survey
give a mean ratio at zero redshift
of $q_{24}\,=\,\log(F_{\rm 24}/F_{\rm 1.4GHz})\,=\,0.84\pm0.28$.
For the 4 active BCDs in our sample, this ratio is exceeded at levels of $\simgt4\sigma$:
$\langle q_{24} \rangle\,=\,2.0\pm0.16$.
Mrk\,33 and II\,Zw\,70 have lower $q_{24}$, $\langle q_{24} \rangle\,=\,1.6$,
still higher than the mean value.
\citet{appleton} find a larger dispersion in $q_{24}$ than in the equivalent 70\,$\mu$m
value, and suggest that there is much more intrinsic dispersion in the IR/radio ratio
at 24\,$\mu$m than at longer wavelengths.
The SED variations of the kind observed in our sample, due either to active/passive 
dichotomy or to metallicity, may be causes of such dispersion.

\section{Age and dust masses \label{sec:age}}

In Sect. \ref{sec:radiofir}, we discussed the hypothesis of \citet{cannon04} 
that the shape of the radio spectrum indicates starburst age.
The IR portion of the spectrum also depends on the age
of the starburst because of the time variation of dust self-absorption
and the UV radiation field \citep[e.g.,][]{takeuchi}.
The effects of metallicity are more difficult to assess, both because of the
small size of our sample and because of the dearth of data for the
``passive'' BCDs.
Nevertheless, we can use our sample to
gain insight into possible dust-formation mechanisms 
and evolution.

The origin of dust in galaxies is directly connected to their
star-formation history (SFH); dust is ultimately produced
by the various byproducts of stellar evolution.
In very young ($\simlt$100\,Myr)
or primordial starbursts, Type II supernovae (SNe)
are the only possible dust production mechanism.
Later, hot massive stars (e.g., Wolf-Rayet) and, still later, cooler
less massive stars (Asymptotic Giant Branch, AGBs) condense dust in their
stellar winds.
Dust production in these various phases has been estimated by
\citet{morgan}; 
at least 100\,Myr are needed to achieve the conditions
for dust nucleation in the stellar evolution
of low- and intermediate-mass stars.

We wish to examine whether the dust masses inferred from our
\dusty\ \,SED models can be attributed to the current starburst,
and whether metallicity plays a role.
To do so, we neglect the contribution of AGBs,
which at 100\,Myr are roughly 1\% of the total dust yield \citep{morgan}.
We also neglect Wolf-Rayet stars \citep{morgan},
because such massive stars are rare, and their dust yield is low,
$\sim10^{-5}$\,\msol/star \citep{marchenko}. 
We therefore use the models of \citet{hhf}, based on  
the dust yields from type II SNe given by \citet{tf}, 
to estimate the amount of dust generated by the current burst.
However, we first need an estimate of SFRs. 
Because of the difficulties described in Sect. \ref{sec:scaling},
it is not clear how we can reliably deduce SFRs for the galaxies
in our sample.
As a ``best guess'', we adopt the thermal radio estimate from
Col. (6) in Table \ref{tab:sfr} for the three galaxies with
approximately flat radio spectral indices (II\,Zw\,40, \sbs, \ngc),
and the 60$\mu$m estimate (Col. 4) for the remaining ones.

At ages of 3--5\,Myr, with the exception of \sbs,
Type II SNe are not capable of producing the
amount of dust inferred from our models;
at 5\,Myr, we would expect $\simgt2\times10^3$\,\msol\ of dust
from the \citet{hhf} models and $\simgt10^2$\,\msol\ from those by
\citet{morgan}. 
These masses are 100--1000 times lower than the $10^5$\,\msol\ we find
from our models of the active BCDs.
Because of the presence of Wolf-Rayet stars (see Sect. \ref{sec:active}),
the present burst in all galaxies except II\,Zw\,70 must be young;
but with the exception of \sbs,
all the galaxies for which we have dust mass estimates have NIR
colors which indicate an evolved stellar population.
Hence, the dust heated by the present burst almost certainly results
from previous episodes of star formation.
For the SFRs of our sample, the dust yields from SNe alone at or 
beyond ages of $\simlt$30\,Myr are more than
sufficient to produce the $\sim10^5$\,\msol\ of dust inferred from our
models \citep{hhf,morgan}.

On the other hand,
the low dust mass we infer for \sbs\ can easily be ascribed to
the current young burst.
Indeed, the dust yields are rather high for the $\sim10^2$\,\msol\ of dust
inferred from our SED.
In this BCD the uncertainty in the SED itself makes the dust
estimate particularly problematic.
However, should our dust mass be correct,
the burst could be extremely young ($\simlt$\,2\,Myr),
even though the radio spectral index is slightly nonthermal ($\sim-0.3$). 
Alternatively, either the dust production of a single Type II SN
could be lower than that assumed by \citet{hhf}, or other
mechanisms could be important such as outflows \citep {lisenfeld}
or dust destruction by SNe shocks \citep[e.g.,][]{dwek80}.
Obviously on the basis of one galaxy it is impossible to ascertain whether 
metallicity, youth, or compact size/density (the ``active/passive'' dichotomy)
are the primary factors in shaping the unusual SED of \sbs.
More MIR and FIR observations of low-metallicity BCDs are needed to
resolve the question.

\section{Metallicity and dust-to-gas ratio \label{sec:metals}}

One of the foundations of models of dust formation and
evolution in galaxies has been the dust-to-gas ratio \dg\
\citep{lisenfeld,edmunds,james02,htk}.
Here we evaluate \dg\ for our sample galaxies with the aim
of reassessing trends of \dg\ with metallicity.

The gas mass in low-metallicity BCDs consists primarily
of atomic hydrogen. Molecular gas in the form of CO
is notoriously difficult to detect in low-metallicity objects,
and of our sample only the highest metallicity
objects, \ngc, He\,2$-$10, Mrk\,33,
have been detected in the CO(1--0) transition
\citep{baas,kobulnicky95,barone,meier01,meier02,bravo}.
\ngc\ has also been detected in a MIR $H_2$ line with ISO,
although II\,Zw\,40 was not \citep{rigopoulou}.
These CO measurements together with HI masses taken
from the literature 
\citep{ks95,vanzee,thuan99a,ks} enable us to calculate total gas masses.
The molecular gas mass in \ngc\ is 0.4\% that of the atomic gas
\citep{meier02},
but more than twice the atomic gas mass in the solar-metallicity
galaxy He\,2$-$10 \citep{kobulnicky95}. 
The lack of CO detections in II\,Zw\,40 and
\sbs\ could imply that the gas in these BCDs is entirely atomic,
or that CO is not a good tracer of H$_2$ at these sub-solar
metallicities \citep[e.g.,][]{taylor98}.
The dust-to-gas ratio \dg\ is then derived by combining our
model results for dust mass, and reporting all quantities to
our assumed distance. 

\begin{figure}
\centering{\includegraphics[angle=0,width=\linewidth,bb=20 145 590 645]{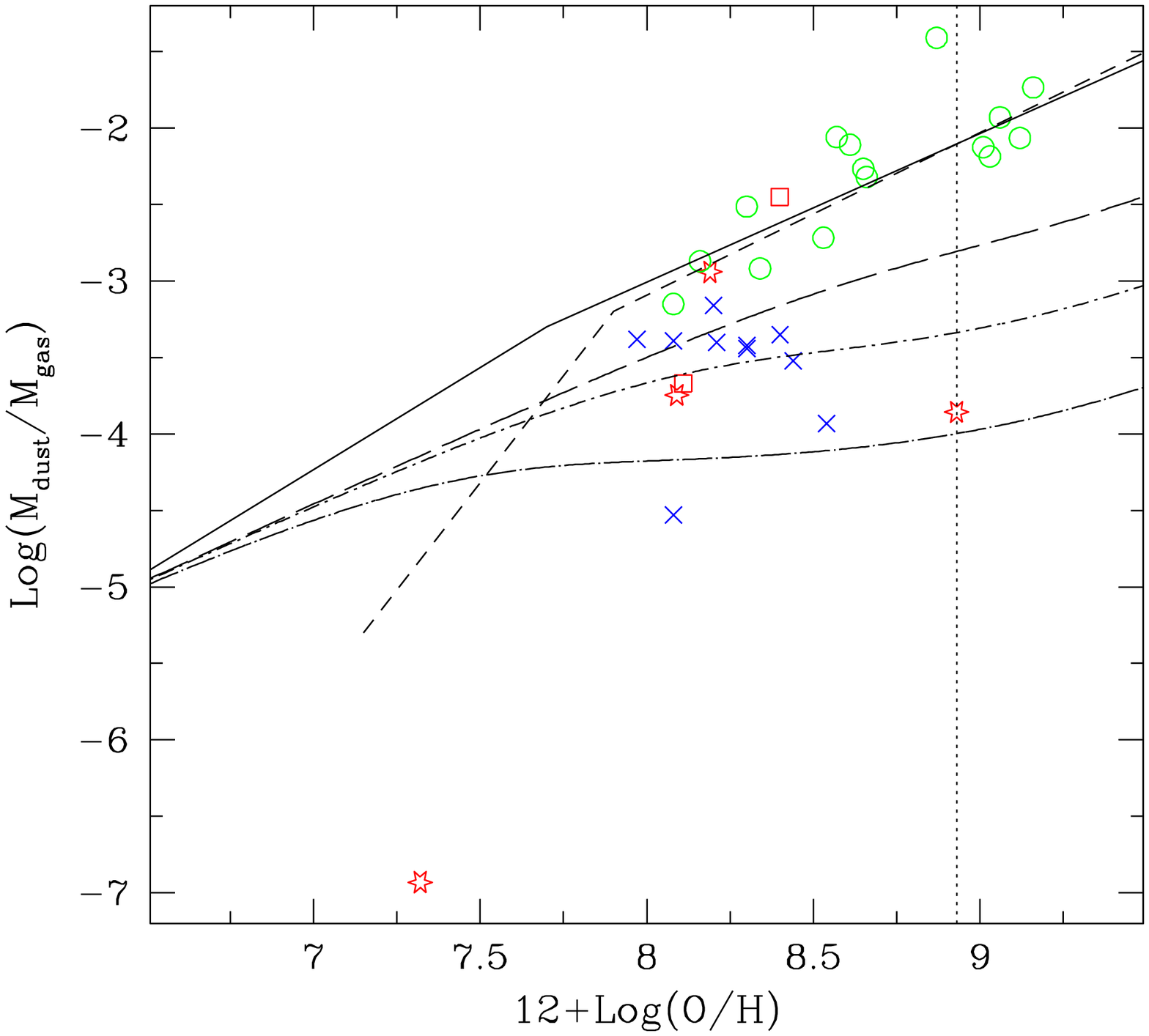}}
\caption{Dust-to-gas mass ratios $\log$\dg\ versus oxygen abundance $12+\log(O/H)$.
The four galaxies for which we have \dusty\ models and thus dust masses
are shown as open stars;
other data from the literature are also plotted
[open circles are from \citet{james02} and $\times$ from \citet{lisenfeld}].
Open squares show data for II\,Zw\,70 \citep{lisenfeld} and
Mrk\,33 \citep{james02}, both of which are in our sample.
The vertical dotted line indicates solar metallicity.
The solid and short-dashed lines are the models by
\citet{james02}, and the long-dashed and dot-dashed lines the models
for various dust destruction efficiencies} by \citet{htk}.
\label{fig:d2g}
\end{figure}

Figure \ref{fig:d2g} shows graphically the known correlation
between \dg\ and oxygen abundance \citep{issa};
our new determinations are shown as open stars, data from
the literature as open circles \citep{james02} and $\times$
\citep{lisenfeld}\footnote{Following \citet{htk}, we have
eliminated their factor of 0.5 in their H{\sc I} mass derivation.};
Seyfert galaxies and galaxies for which there are no IRAS
detections are not included.
We have also indicated with open squares the two galaxies in our
sample which follow the standard starburst template,
and for which we did not have sufficient data to perform a \dusty\
fit:
II\,Zw\,70 \citep{lisenfeld} and Mrk\,33 \citep{james02}.
\ngc\ appears in the \citet{james02} sample with $\log$\dg\,=\,$-2.92$,
in very good agreement with our value $\log$\dg\,=\,$-2.94$.

Model predictions for the \dg-metallicity correlation are also shown 
in Fig. \ref{fig:d2g} \citep{edmunds,htk}.
The \citet{htk} models, shown as long-dashed and dot-dashed
curves, vary because of the dust destruction
efficiency which begins to be effective when oxygen abundance $12+\log(O/H)$
reaches $\sim8$.
Their models neglect outflow, and are complementary
to \citet{lisenfeld} who instead attribute the variation
in \dg\ to variations in the mass outflow rate.
Both the \citet{lisenfeld} and \citet{htk} models predict a non-linear
trend for \dg$\,-\,O/H$, unlike \citet{edmunds} and \citet{james02} who
predict a linear one,
proposing that the interstellar dust mass is an approximately constant
fraction of the ISM metal abundance.
A solid line indicates their prediction for dust production by SNe, 
and the short-dashed one
represents that by only evolved low- and intermediate-mass stars.
It is apparent from Fig. \ref{fig:d2g} that none of the models
reproduces the trends of \dg$\,-\,O/H$ observed in our and other BCD
samples.
\citet{james02} argue that \citet{lisenfeld} underestimated the cool-dust
component because of the unavailability of sub-mm fluxes; this
would push the $\times$'s upwards in Fig. \ref{fig:d2g}.
Nevertheless, our estimates of \dg\ which take into account the cool dust in a 
self-consistent way are equally deviant from a linear model.
In addition to outflow and dust destruction mechanisms,
intermittent SFH and age probably play important roles in
governing \dg\ \citep{htk}.
However, without significantly more data at low metallicities ($12+\log(O/H)\simlt8$),
it will be difficult to decide between the linear/non-linear alternatives.
As pointed out by \citet{james02}, dust detections in low-metallicity
star forming galaxies are badly needed.

\section{Implications for photometric redshifts \label{sec:highz}}

The determination of the spectroscopic redshift of distant
submm/mm sources has been extremely challenging due to their
very weak and uncertain optical counterparts. However,
the problem of the redshift determination has been partially
solved by using their far-IR--to--radio SED, compared to local
templates, to obtain a photometric redshift \citep{yun,hughes}.
This method has been tested to successfully recover the spectroscopic
redshifts with a reasonable accuracy ($\rm z_{phot}-z_{spec}< 0.3$,
at $\rm 1<z<3$). Generally, even the simple submm-to-radio
spectral index ($\rm \alpha _{1.4GHz}^{350GHz}$) provides a good
approximation of the actual redshift \citep{carilli}.
These radio-submm photometric
redshifts have been calibrated by using templates of local, massive and
chemically evolved starburst galaxies (e.g. M82, Arp220; here we focus
on the non-AGN cases), which are characterized by far-IR bumps peaking
at $\sim$60--100$\mu$m and a radio tail dominated by synchrotron emission.
Such templates are certainly adequate for the submm-mm sources at
${\rm z}\sim2.5$ identified by SCUBA and MAMBO 
\citep{chapman,dannerbauer}, which are massive and metal-rich
\citep[e.g.,][]{tecza,genzel}.
However, the starburst galaxies which will be detected at very high
redshift (${\rm z}\ge6$) by the forthcoming submm facilities
(and ALMA in particular) will probably be characterized by
compact star forming regions and low metallicities, similar
to some of the local BCDs discussed in this paper. As consequence,
the SED of primordial galaxies at ${\rm z}>6$ will likely be characterized by
warmer infrared bumps and thermal radio emission. 
The use of ``standard'' (M82-like) SEDs applied to
primordial galaxies with SEDs similar to the BCDs in our
sample would yield deceptively high, significantly
overestimated photometric redshifts (up to a factor of 2).

Another related factor which is likely to affect the SED of starburst
galaxies at ${\rm z}>6$ is that at such early epochs (age of the
universe $<$1\,Gyr) AGB stars have not yet evolved to produce dust.
At ${\rm z}>6$ the only viable mechanism for dust production
is through Type II supernovae \citep{maiolino04}. SN dust
has different properties than ``standard'' dust
(smaller grains), resulting in warmer SEDs. 
As discussed in Sect. \ref{sec:metals}, a 
SN component of dust is also likely to contribute to
the warm SED of the BCD \sbs\ \citep{hhf}.
Photometric redshifts for primordial galaxies calibrated
by using the SEDs of local BCDs will be presented in a
forthcoming paper.

\section{Conclusions}

We have used new and archival data to
derive the optical-to-radio spectral energy distributions
for a sample of seven low-metallicity BCDs.
Such galaxies are regarded as the best local laboratories
for high redshift, primordial galaxies, forming stars for
the first time.  In particular, their
infrared and radio emission provide important constraints on the early
stages of star formation and on the associated dust properties.

The SEDs of BCDs have revealed a wealth of information as well as
unexpected, important properties.
The main results can be summarized as follows:

\begin{enumerate}

\item The BCD SEDs deviate significantly from the
standard templates of ``classical'', evolved and massive starburst
galaxies
in several ways.
In particular:  ({\it i})~the radio spectrum is generally flatter,
indicative of thermal emission;
({\it ii})~the shape of the
infrared bump generally implies warmer dust temperatures
(with \sbs\ being an extreme case in this respect);
({\it iii})~the mid-IR spectra show little or no Aromatic Features
in Emission (e.g., PAHs).

\item The different spectral properties of BCDs relative
to ``normal'' starburst galaxies also yield 
strong deviations from the ``scaling laws''. In particular,
BCDs deviate substantially from the radio/FIR and MIR/FIR
relations. One of the consequences of such deviations is that
the SFR inferred from any of these indicators
(radio, FIR, MIR) may be wrong by large factors.

\item The origin of the different SEDs of BCDs with respect to
``classical'' starburst galaxies is not clear yet. Very likely
the very compact and dense nature of the star forming regions
in some BCDs (the ``active'' ones) plays a major role, and
is probably responsible for heating the dust to higher temperatures
and for the absence of AFEs. Age may also be important, particularly
in the dominance of the thermal emission over synchrotron
in the radio.
A young age is also probably related to the compactness of the star
forming regions discussed above. The different origin of dust in the
early stages of star formation (supernovae vs. evolved stars),
resulting in different dust properties (typically smaller grains), is
another factor which is likely to play an important role in the shape
of the IR SEDs. The influence of metallicity is not clear, and additional
data are definitely required to further investigate its effects on
the SEDs of star-forming galaxies.

\item We have modelled the IR SEDs by means of a dust radiative
transfer code, which has allowed us to derive the
infrared luminosities, dust radial distribution, masses and temperatures.
An important result is that, with
the notable exception of \sbs, the observed dust mass
cannot be accounted for by the dust produced during the short
episode of star formation traced by the still ongoing burst; most
of the dust must have been produced by supernovae or AGB stars from
previous episodes of star formation. 
Another important result
is that the dust-to-gas ratio is not a linear function of the
metallicity, nor does the observed relation follow the prediction of
any of the more sophisticated models. 
The latter result suggests that
the dust production mechanisms in different metallicity environments
need to be reconsidered.

\item Finally, the significant deviations of the BCD SEDs with
respect to ``classical'' evolved and massive starburst galaxies
prompt for a revision of the photometric redshift techniques
to be applied to the primordial, very
high redshift galaxies, expected to be detected by facilities
such as ALMA and SKA.

\end{enumerate}

\begin{acknowledgements}
We warmly thank F. Bertoldi and K. Menten
who made available their MAMBO measurement of \izw.
We are also grateful to an astute referee whose comments helped
improve the paper.
This research has made use of the NASA/IPAC Extragalactic Database (NED) which
is operated by the Jet Propulsion Laboratory, California Institute of
Technology, under contract with the National Aeronautics and Space
Administration.
We also acknowledge the 
Canadian Astronomy Data Centre, which is operated by the
Dominion Astrophysical Observatory for the National Research Council of
Canada's Herzberg Institute of Astrophysics.
\end{acknowledgements}


\appendix
\section{SED data and references \label{app:data}}

The fluxes used in the SEDs together with their references are given in
Tables A1 through A7 for He\,2$-$10, II\,Zw\,40, II\,Zw\,70, \izw,
Mrk\,33, \ngc,  and \sbs, respectively.
Wavelengths in the tables are given as $\mu$m up to the ``Radio'' region,
where they are given in cm.
The values reported here have not been corrected for Galactic extinction.
The cross-identification of the data references follow the
individual data tables.


\begin{figure*}[t]
\begin{center}
\includegraphics[angle=0,width=0.32\linewidth,bb=1 221 290 765]{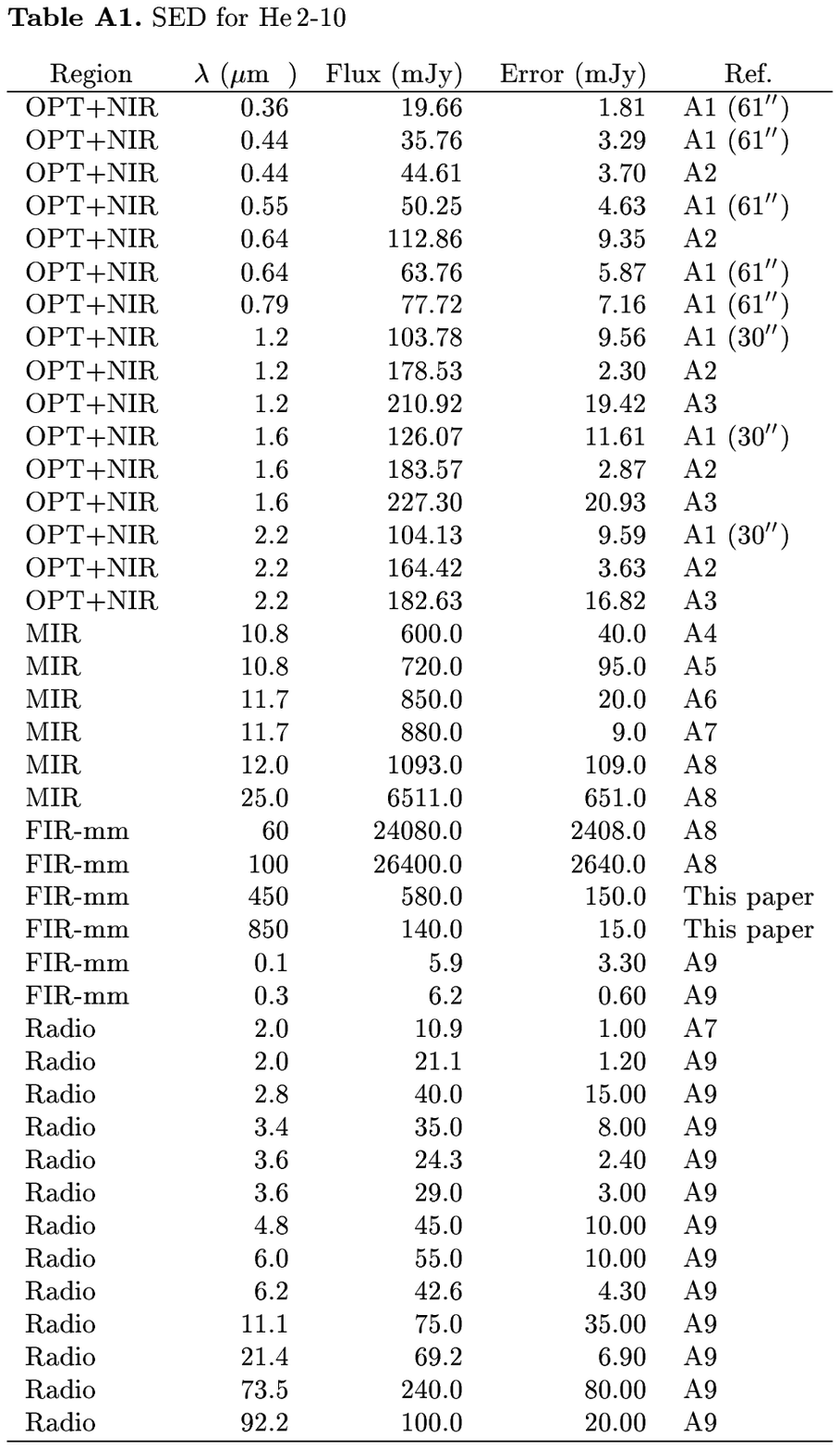}
\hspace{1.0cm}
\includegraphics[angle=0,width=0.32\linewidth,bb=1 221 290 765]{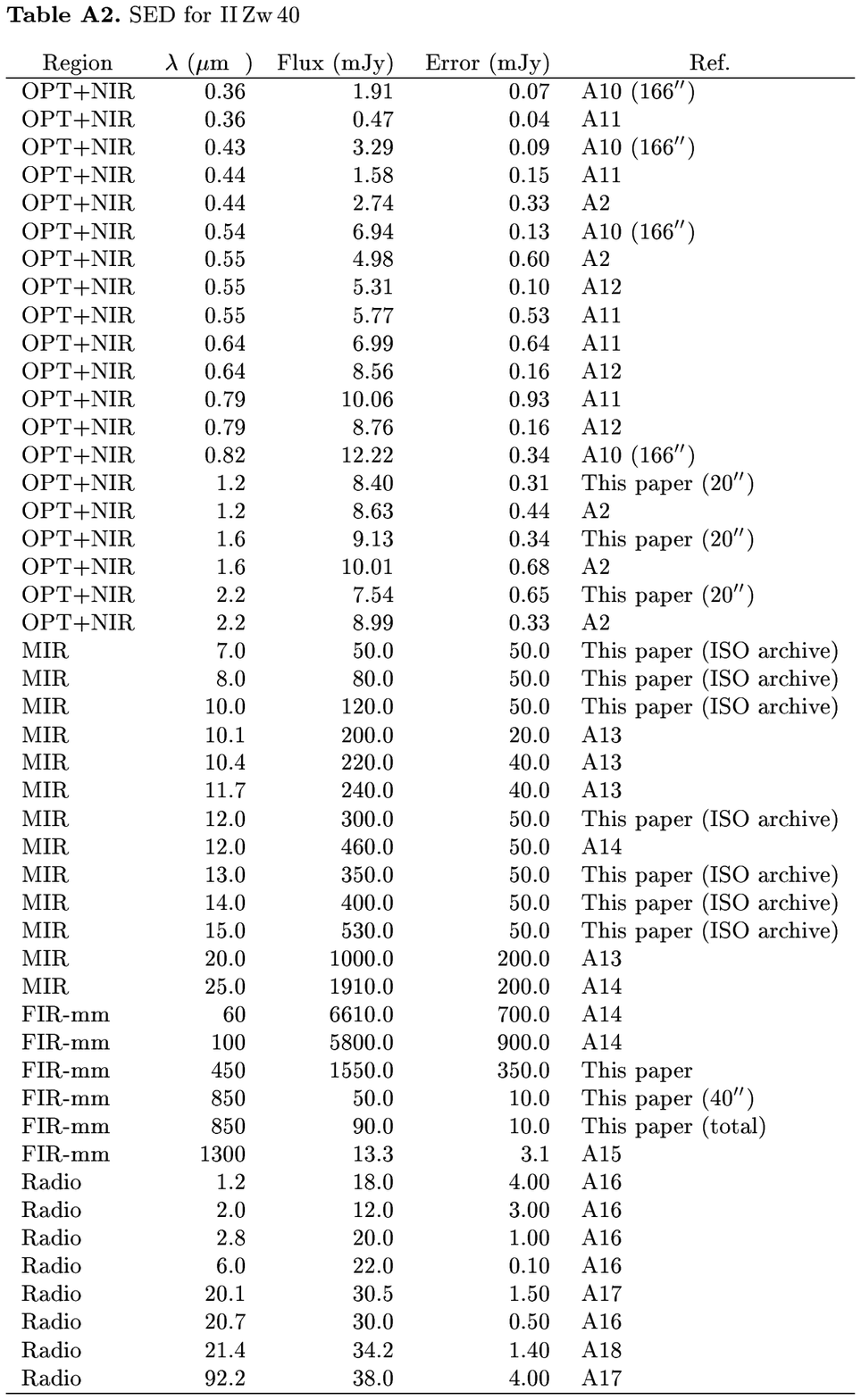}
\end{center}
\vspace{0.5cm}
\begin{center}
\includegraphics[angle=0,width=0.32\linewidth,bb=1 407 290 765]{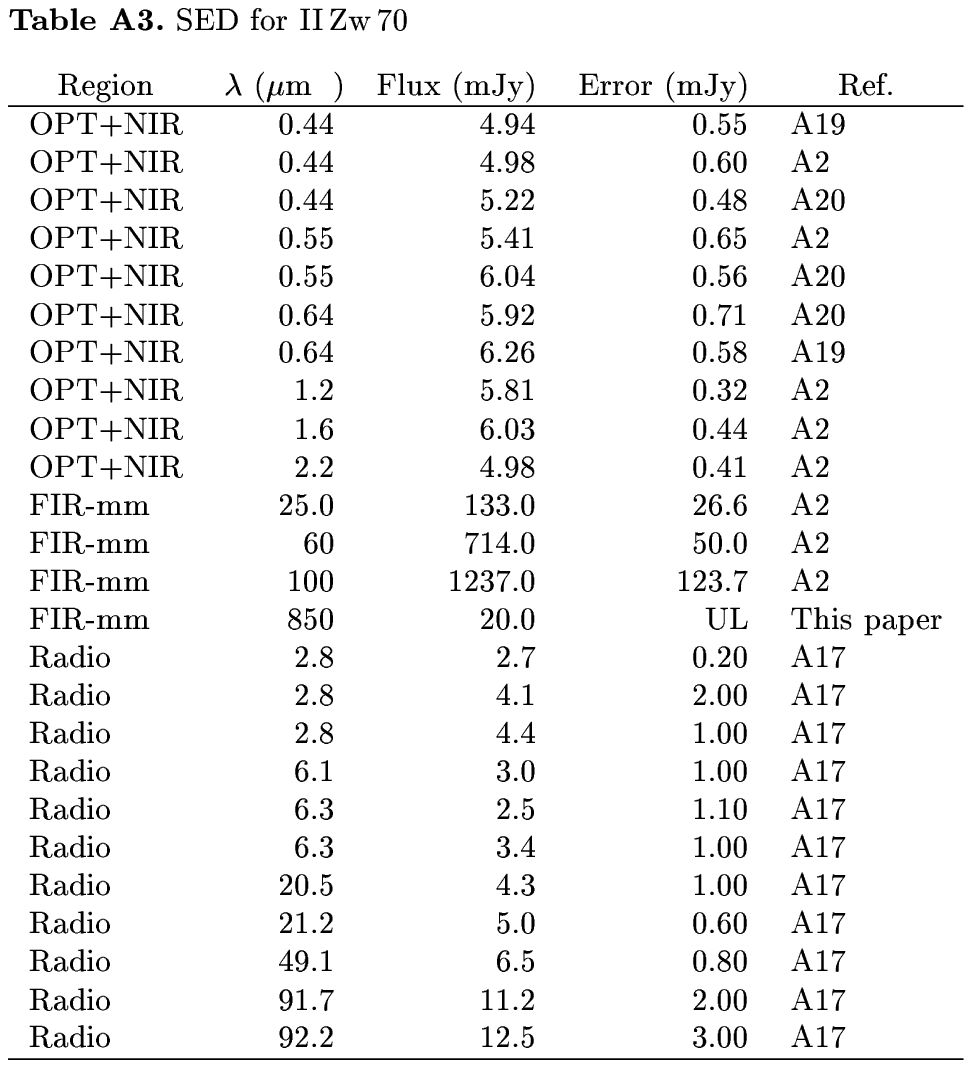}
\hspace{1.0cm}
\includegraphics[angle=0,width=0.32\linewidth,bb=1 407 290 765]{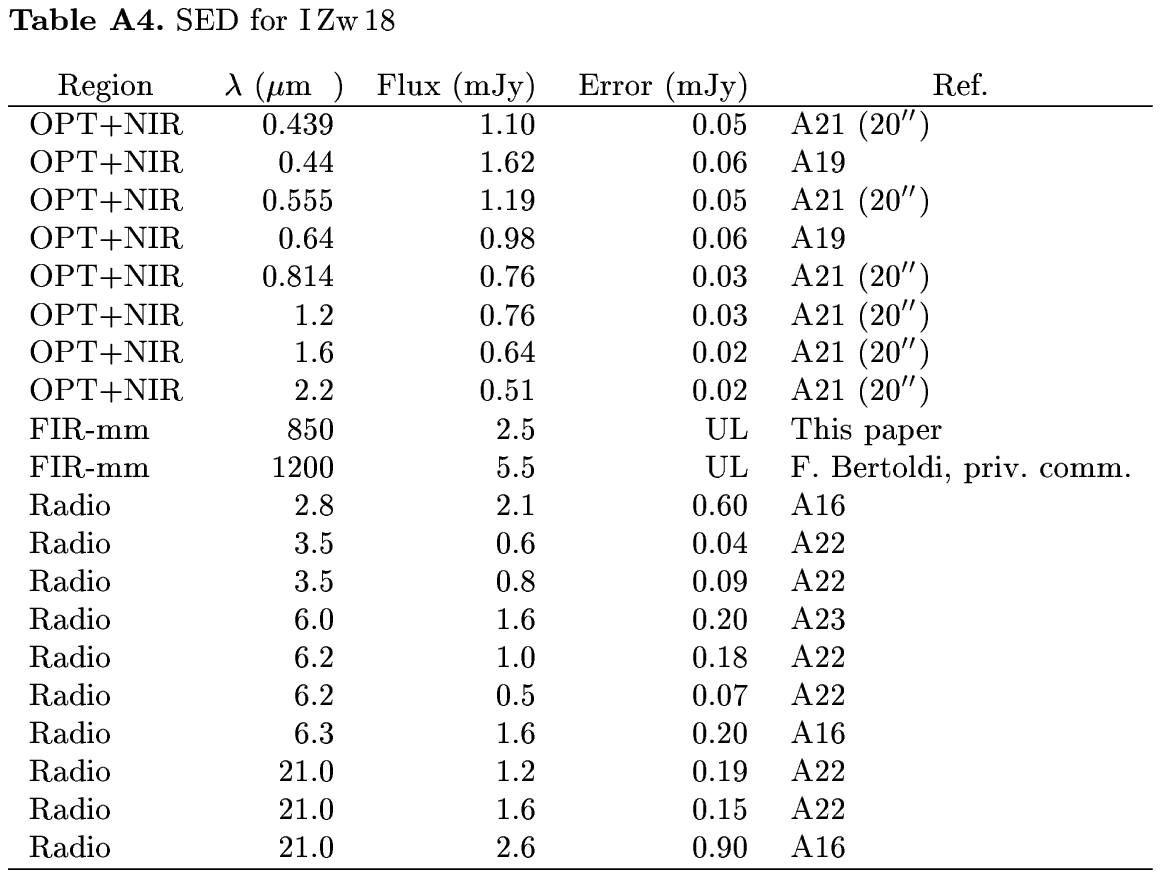}
\end{center}
\end{figure*}
\begin{figure*}[t]
\begin{center}
\includegraphics[angle=0,width=0.32\linewidth,bb=1 221 290 765]{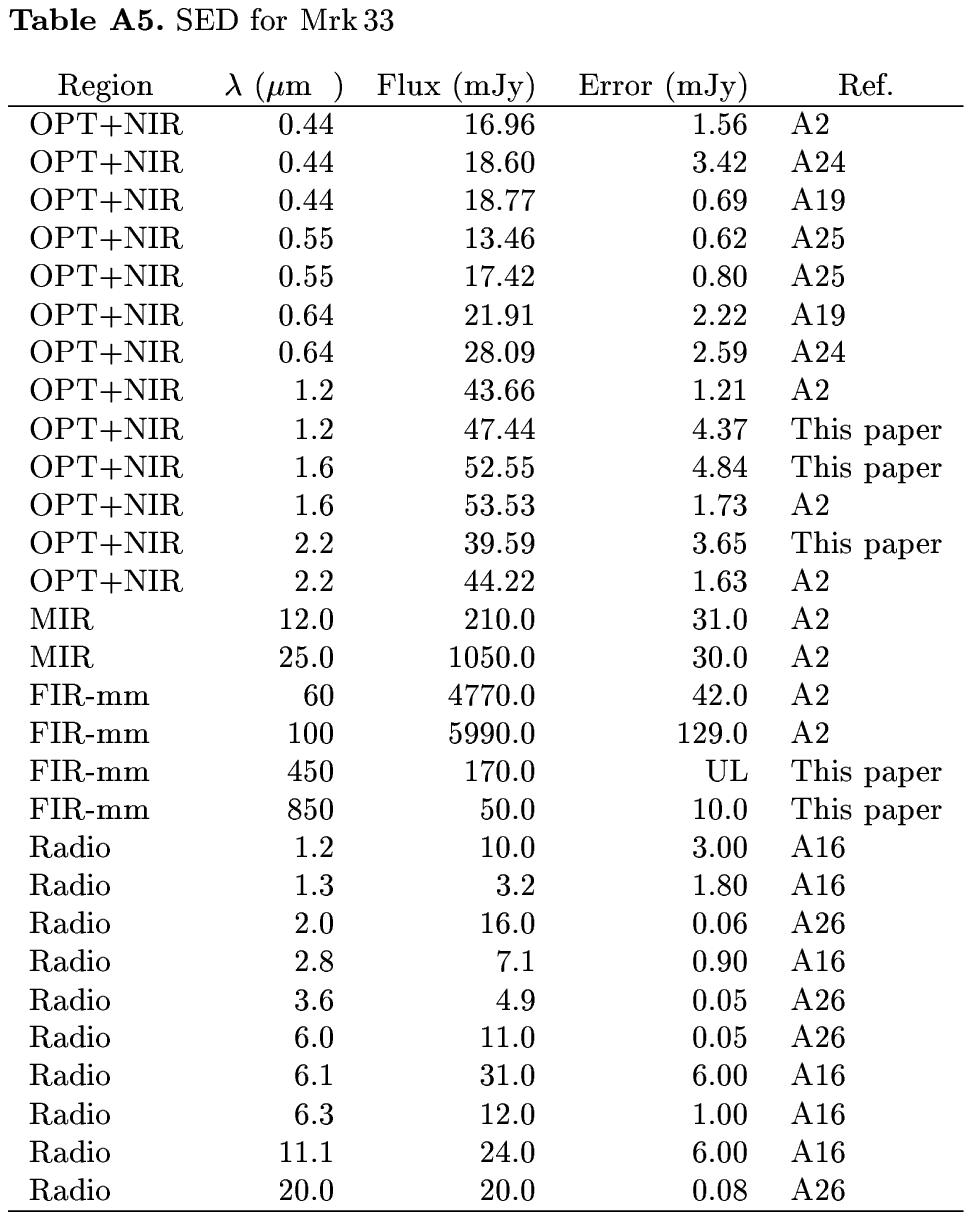}
\hspace{1.0cm}
\includegraphics[angle=0,width=0.32\linewidth,bb=1 221 290 765]{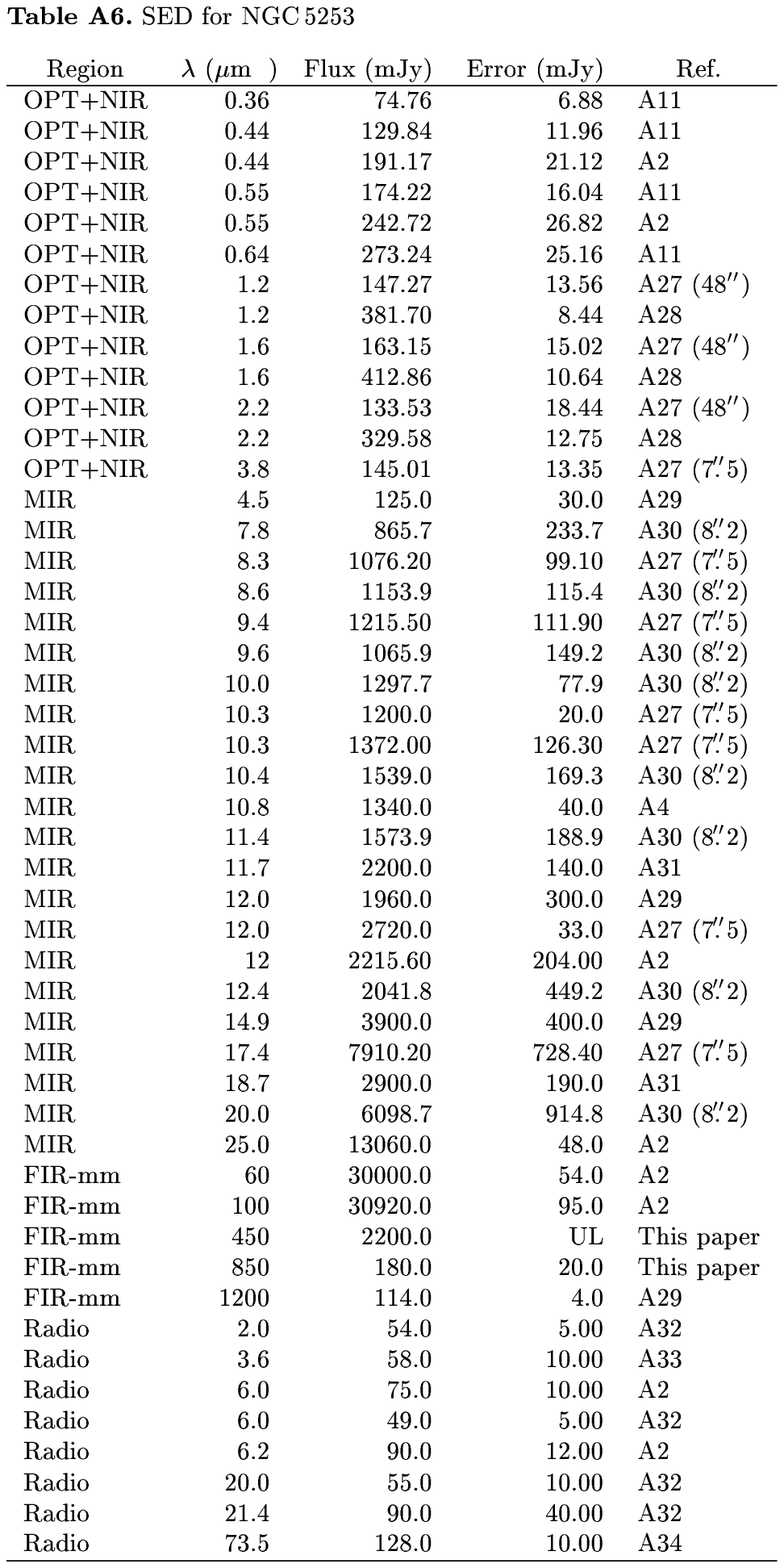}
\end{center}
\end{figure*}

\begin{figure*}[t]
\begin{flushleft}
\hspace{2.5cm}
\includegraphics[angle=0,width=0.32\linewidth,bb=1 95 290 560]{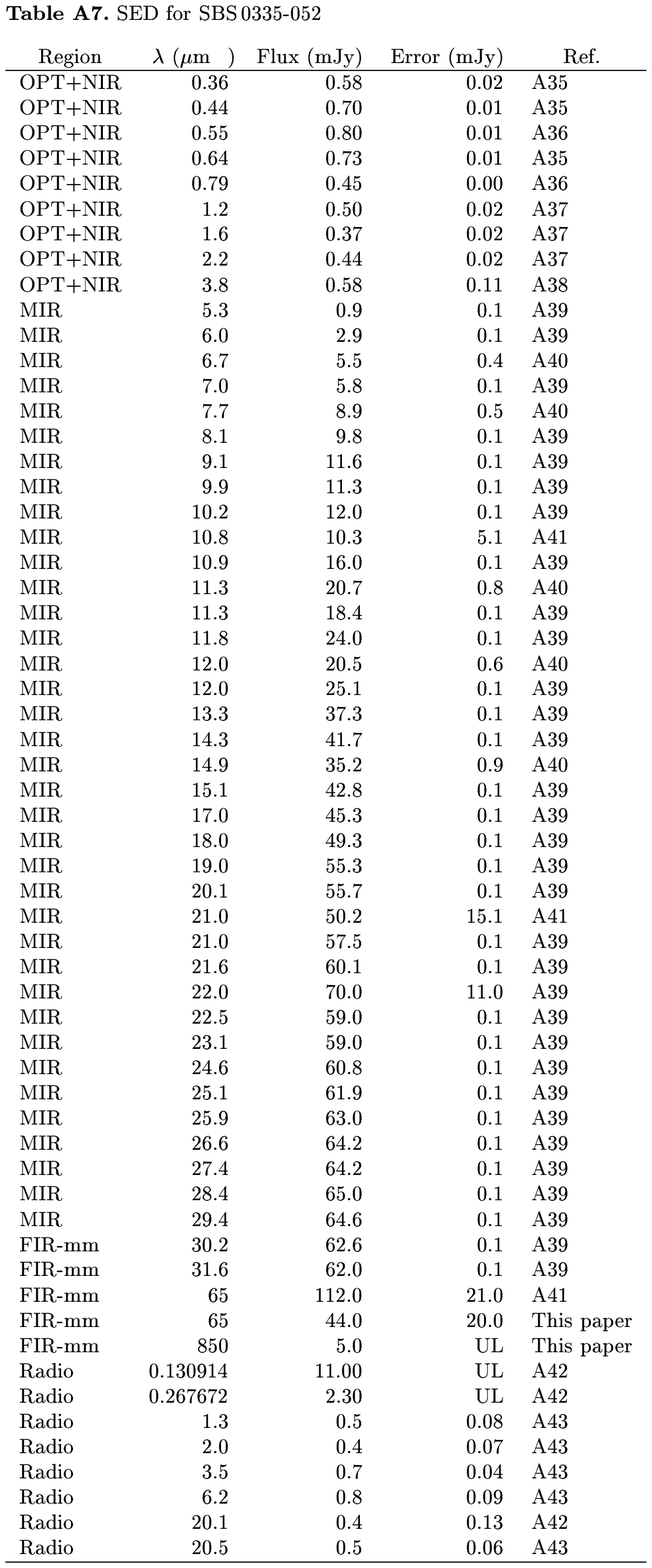} 
\end{flushleft}
\end{figure*}


\newcounter{aaa}

\bigskip
\bigskip

\noindent
{\large\sffamily\bfseries  Data references}
\begin{list}
{[A\arabic{aaa}]}{\usecounter{aaa}}
\setlength{\itemsep}{0pt}
\setlength{\parskip}{0pt}
\item
Johansson, I.\ 1987, A\&A, 182, 179 
\item
NASA Extragalactic Database (NED)
\item
Noeske, K.~G., Papaderos, P., Cair{\'o}s, L.~M., 
\& Fricke, K.~J.\ 2003, A\&A, 410, 481 
\item
Telesco, C.~M., Dressel, L.~L., \& Wolstencroft, R.~D.\ 1993, ApJ, 414, 
120 
\item
Vacca, W.~D., Johnson, K.~E., \& Conti, P.~S.\ 2002, AJ, 123, 772 
\item
Sauvage, M., 
Thuan, T.~X., \& Lagage, P.~O.\ 1997, A\&A, 325, 98 
\item
Beck, S.~C., 
Turner, J.~L., \& Gorjian, V.\ 2001, AJ, 122, 1365 
\item
ISSA/OASIS
\item
Kobulnicky, 
H.~A.~\& Johnson, K.~E.\ 1999, ApJ, 527, 154
\item
Marlowe, A.~T., Meurer, G.~R., Heckman, T.~M., \& Schommer, R.\ 1997, 
ApJS, 112
\item
Heisler, C.~A.~\& Vader, J.~P.\ 1994, AJ, 107, 35 
\item
Telles, E.~\& Terlevich, R.\ 1997, MNRAS, 286, 183 
\item
Beck, S.~C., Turner, 
J.~L., Langland-Shula, L.~E., Meier, D.~S., Crosthwaite, L.~P., \& Gorjian, 
V.\ 2002, AJ, 124, 2516 
\item
Vader, J.~P., Frogel, J.~A., Terndrup, D.~M., \& Heisler, C.~A.\ 1993, AJ, 
106, 1743 
\item
Albrecht, M., Chini, 
R., Kr{\" u}gel, E., M{\" u}ller, S.~A.~H., \& Lemke, R.\ 2004, A\&A, 414, 
141 
\item
Klein, U., Weiland, H., \& Brinks, E.\ 1991, A\&A, 246, 323 
\item
Deeg, H., Brinks, E., 
Duric, N., Klein, U., \& Skillman, E.\ 1993, ApJ, 410, 626 
\item
Condon, J. J., Cotton, W. D., Greisen, E. W., Yin, Q. F., Perley, 
R. A., Taylor, G. B., \& Broderick, J. J. 1998, AJ, 115, 1693: 
National Radio Astronomy Observatory Very Large Array Sky Survey (NVSS)
\item
Gil de Paz, A., Madore, B.~F., \& Pevunova, O.\ 2003, ApJS, 147, 29 
\item
Cox, A.~L., Sparke, L.~S., Watson, A.~M., \& van Moorsel, G.\ 2001, AJ, 
121, 692 
\item
Hunt, L.~K., Thuan, T.~X., \& Izotov, Y.~I.\ 2003, ApJ, 588, 281 
\item
Hunt, L.~K., Dyer,~K., \& Thuan, T.~X. 2004, in preparation
\item
Lequeux, J.~\& Viallefond, F.\ 1980, A\&A, 91, 269 
\item
Summers, L.~K., Stevens, I.~R., \& Strickland, D.~K.\ 2001, MNRAS, 327, 
385 
\item
Huchra, J.~P. 1977, ApJS, 35, 171
\item 
Beck, S.~C., Turner, J.~L., \& Kovo, O.\ 2000, AJ, 120, 244 
\item
Moorwood, A.~F.~M.~\& Glass, I.~S.\ 1982, A\&A, 115, 84 
\item
Jarrett, T.~H., 
Chester, T., Cutri, R., Schneider, S.~E., \& Huchra, J.~P.\ 2003, AJ, 125, 
525 
\item
Vanzi, L.~\& Sauvage, M.\ 2004, A\&A, 415, 509 
\item
Frogel, J.~F., Elias, J.~H., \& Phillips, M.~M.\ 1982, ApJ, 260, 70 
\item
Gorjian, V., Turner, J.~L., \& Beck, S.~C.\ 2001, ApJL, 554, L29 
\item
Beck, S.~C., Turner, 
J.~L., Ho, P.~T.~P., Lacy, J.~H., \& Kelly, D.~M.\ 1996, ApJ, 457, 610 
\item
Mohan, 
N.~R., Anantharamaiah, K.~R., \& Goss, W.~M.\ 2001, ApJ, 557, 659 
\item
Schmitt, H.~R., Kinney, A.~L., 
Calzetti, D., \& Storchi Bergmann, T.\ 1997, AJ, 114, 592 
\item
Papaderos, P., 
Izotov, Y.~I., Fricke, K.~J., Thuan, T.~X., \& Guseva, N.~G.\ 1998, A\&A, 
338, 43 
\item
Thuan, T.~X., Izotov, Y.~I., \& Lipovetsky, V.~A.\ 1997, ApJ, 477, 661 
\item
Vanzi, L., Hunt, L.~K., Thuan, T.~X., \& Izotov, Y.~I.\ 2000, A\&A, 363, 493 
\item
Hunt, L.~K., Vanzi, L., \& Thuan, T.~X.\ 2001, A\&A, 377, 66 
\item
Houck, J.~R., et al.\ 2004, ApJS, 154, 211 
\item
Thuan, T.~X., Sauvage, M., \& Madden, S.\ 1999, ApJ, 516, 783 
\item
Plante, S.~\& Sauvage, M.\ 2002, AJ, 124, 1995 
\item
Dale, D.~A., Helou, G., 
Neugebauer, G., Soifer, B.~T., Frayer, D.~T., \& Condon, J.~J.\ 2001, AJ, 
122, 1736 
\item
Hunt, 
L.~K., Dyer, K.~K., Thuan, T.~X., \& Ulvestad, J.~S.\ 2004, ApJ, 606, 853 
\end{list}

\section{Dust mass from \dusty \label{app:equations}}

Here we set out the equations used to infer the dust
mass from the output given by \dusty.
Assuming that the dust is distributed in a spherical shell with
a power-law radial density profile $\rho(r)=C_1 r^{-p}$,
the dust mass can be written as:
\begin{eqnarray}
M_{\rm dust} & = & \int_{R_{\rm in}}^{R_{\rm out}} 4\pi r^2 C_1 r^{-p} dr \nonumber \\
& = & 4 \pi C_1
                 \left\{ 
                 \begin{array}{lc} 
                         \frac{R_{\rm out}^{3-p} - R_{\rm in}^{3-p}}{3-p} & p \neq 3   \\
			 \ln\frac{R_{\rm out}}{R_{\rm in}}          & p = 3 
                 \end{array}
                 \right. 
\label{eqn:mdust}
\end{eqnarray}
\noindent
where $C_1$ is a constant, and $R_{\rm in}$ and $R_{\rm out}$ are the inner and outer shell radii.

\dusty\ provides the value of $\tau_\nu$, 
the optical depth at frequency $\nu$, which can be expressed as:
\begin{eqnarray}
\tau_\nu & = & \int_{R_{\rm in}}^{R_{\rm out}} {\cal X_\nu} (r) dr \nonumber \\
& = & \int_{R_{\rm in}}^{R_{\rm out}} dr\, \int_{a_1}^{a_2} da\, \pi a^2 \left[ \sum_i Q_{i,\nu}(a)\, n_i (a,r) \right] 
\label{eqn:taudust}
\end{eqnarray}
\noindent
where $i$ refers to the grain chemical composition type.
$\cal X_\nu$ is the absorption coefficient per unit length;
$Q_{i,\nu}(a)$ the extinction efficiency factor (ratio of optical to geometrical cross section)
for type $i$ grains with size (radius) $a$;
$n_i (a,r)$ the number density distribution in radius for grains of type $i$
at radius $r$ of size (radius) $a$; 
and $\pi a^2$ the geometrical cross section of a grain with size (radius) $a$. 
$a_1$ and $a_2$ are the minimum and maximum grain sizes, respectively.
$n_i (a,r)$ can be written as two functions which separate the mass distribution
radial dependence from the grain size:
$n_i (a,r)\,=\,N_i(r) {\cal F}(a)$.
Assuming the grain sizes are distributed as a power law,
${\cal F}(a)\,=\,A a^{-q}$, where $A$ is such that
$\int_{a_1}^{a_2} {\cal F}(a)\,da\,=\,1$.
$N_i(r)$ is the number density of grain type $i$ at radius $r$.

We can now write $\rho_i$ the mass density at radius $r$ of grain type $i$ as:
\begin{eqnarray}
\rho_i(r) & = & \int_{a_1}^{a_2} \delta_i \left(\frac{4}{3} \pi a^3 \right) n_i(a,r) \,da \nonumber \\
& = & \frac{4}{3} \pi \delta_i N_i(r) A B 
\end{eqnarray}
where $B\,\equiv\,\int_{a_1}^{a_2} a^{3-q}\,da$ and
$\delta_i$ is the mass density of the material composing grains of type $i$.
But $\rho_i(r)$ can also be expressed as $\rho_i(r)\,=\,f_i \rho(r)\,=\,f_i C_1 r^{-p}$,
where $f_i$ is the fractional mass density abundance for grains of chemical type $i$.
We can then solve for $N_i(r)$ in terms of the constant $C_1$:
\begin{eqnarray}
N_i(r) & = & \frac{f_i C_1 r^{-p}}{\frac{4}{3} \pi \delta_i A B} \nonumber \\
& = & C_1 D_i r^{-p}
\end{eqnarray}
\noindent
where $D_i\,\equiv\,\frac{f_i}{\frac{4}{3}\pi \delta_i A B}$, and depends only
on the grain properties.
Equation \ref{eqn:taudust} can then be rewritten in terms of $D_i$ and $C_1$:
\begin{eqnarray}
\tau_\nu & = & \int_{R_{\rm in}}^{R_{\rm out}} dr\, \sum_i N_i(r)
\int_{a_1}^{a_2} da\, \pi a^2 Q_{i,\nu}(a) {\cal F}(a) \\
& = & \int_{R_{\rm in}}^{R_{\rm out}} dr\, C_1 \sum_i D_i r^{-p} \nonumber \\
&& \quad\quad\quad\quad\quad\quad\quad\quad 
\int_{a_1}^{a_2}  da\, (\pi a^2) Q_{i,\nu}(a) A a^{-q} \\
& = & \pi C_1 A
                 \left\{ 
                 \begin{array}{lc} 
                         \frac{R_{\rm out}^{1-p} - R_{\rm in}^{1-p}}{1-p} & p \neq 1   \\
			 \ln\frac{R_{\rm out}}{R_{\rm in}}          & p = 1 
                 \end{array}
                 \right\} \nonumber \\
&& \quad\quad\quad\quad\quad\quad \left[ \sum_i D_i \langle Q_{i,\nu}(a)\rangle \int_{a_1}^{a_2} a^{2-q} da \right]
\end{eqnarray}
Here $\langle Q_{i,\nu}(a)\rangle$ is a mean value, since it has been
taken out of the integral.

The quantities
$R_{\rm in}$, $R_{\rm out}$, $p$, $a_1$, $a_2$, $q$ are input; hence we can calculate the constants
$A$ and $B$.
\dusty\ requires the specification
of the fractional number abundances of the grains $F_i\,=\,N_i/\sum_i N_i$, 
rather than the fractional mass density abundances $f_i$.
However, the two quantities are related through: 
\begin{equation}
f_i  =  \rho_i/\rho \nonumber   =  \frac{\delta_i N_i}{\sum_i \delta_i N_i} 
  =  \frac{\delta_i F_i}{\sum_i \delta_i F_i}
\end{equation}
Therefore, 
with the density $\delta_i$ for each grain type $i$, $D_i$ can be
calculated.
Assigning a mean extinction efficiency factor $\langle Q_i \rangle$
to each grain type,  
we can finally relate $\tau_\nu$ to the constant $C_1$, necessary to derive
the dust mass as given in Equation \ref{eqn:mdust}.
In our calculations,
we have assumed the following physical properties for the grains:
$\delta_{\rm AMC}\,=\,1.81$\,g\,cm$^{-3}$ \citep{kim};
$\delta_{\rm silicate}\,=\,3.3$\,g\,cm$^{-3}$ \citep{dl};
$\delta_{\rm graphite}\,=\,2.26$\,g\,cm$^{-3}$ \citep{dl};
and $\langle Q_i \rangle\,=\,2$ for all grain types \citep[see][]{spitzer}.


\end{document}